\documentclass[%
reprint,
superscriptaddress,
preprintnumbers,
amsmath,amssymb,
aps,
pra,
floatfix,
]{revtex4-2}

\usepackage{graphicx}% Include figure files
\usepackage{dcolumn}% Align table columns on decimal point
\usepackage{bm}% bold math
\usepackage[compat=1.1.0]{tikz-feynhand}
\usepackage{physics}
\usepackage{slashed}
\usepackage{hyperref}
\usepackage{subfigure}
\usepackage{booktabs}
\usepackage{dsfont}
\usepackage{nicematrix}
\usepackage[normalem]{ulem}

\newcommand\numberthis{\addtocounter{equation}{1}\tag{\theequation}}

\makeatletter
\renewcommand\subsubsection{\@startsection{subsubsection}{3}{\z@}%
                                     {-3.25ex\@plus -1ex \@minus -.2ex}%
                                     {1.5ex \@plus .2ex}%
                                     {\normalfont\normalsize\bfseries\boldmath}}
\makeatother

\begin{document}

% ---------------------------------------------------------------------------- %
% ----------------------------------- Misc. ---------------------------------- %
% ---------------------------------------------------------------------------- %

\preprint{ADP-25-7/T1269}

\title{Nucleon resonance structure to 2 GeV and the nature of the Roper}

\author{S.~Owa}
\affiliation{CSSM and ARC Centre of Excellence for Dark Matter Particle Physics, Department of Physics, University of Adelaide, South Australia 5005, Australia}
\author{D.~B.~Leinweber}
\affiliation{CSSM, Department of Physics, University of Adelaide, South Australia 5005, Australia}
\author{A.~W.~Thomas}
\affiliation{CSSM and ARC Centre of Excellence for Dark Matter Particle Physics, Department of Physics, University of Adelaide, South Australia 5005, Australia}

\date{\today}% It is always \today, today,

% ---------------------------------------------------------------------------- %
% --------------------------------- Abstract --------------------------------- %
% ---------------------------------------------------------------------------- %
\begin{abstract}

The study of the spectrum of excited states of the nucleon is vital to our understanding of how QCD is realized in the baryon spectrum. Using a simultaneous analysis of the pion nucleon scattering data up to 2 GeV as well as the results of lattice QCD calculations, we obtain new insight into the nature of the Roper resonance as well as the excited states around 1.9 GeV.
\end{abstract}
%\end{abstract}

\maketitle

% ---------------------------------------------------------------------------- %
% ------------------------------- Introduction ------------------------------- %
% ---------------------------------------------------------------------------- %
\section{Introduction}

The study of the baryon spectrum is expected to lead to a deeper understanding of QCD. It is now apparent that the interpretation of the baryon spectrum is made particularly difficult by the appearance of molecular and dynamically generated states in addition to the classical quark model states. While there is now a consensus that the $\Lambda(1405)$ is predominantly a $\bar{K} N$ bound state~\cite{Hall:2014uca,Hall:2016kou,Oller:2000fj,Kaiser:1995eg,Veit:1984jr,Dalitz:1959dn,Molina:2016LambdaLQCD}, the nature of the first positive excited parity state of the nucleon, the Roper resonance, $N(1440)$, is still controversial~\cite{Leinweber:2015kyz,Kiratidis:2016hda,Wu:2017qve,Burkert:2017djo,Segovia:2015hra,Liu:2017Roper}. Even the first negative parity excitation of the nucleon, the $N(1535)$, has been the subject of multiple interpretations~\cite{Kaiser:1995cy,Bruns:2010sv,Sekihara:2015gvw,Liu:2015ktc,Abell:2023nex,Stokes:2020ElasticFormFactor}.

A great deal of effort has been put into understanding the nature of the Roper resonance through various interpretations and methods~\cite{Weber:1990CQM,Diaz:2006CQM,Golli:2008CQM,Golli:2009CQM,Meissner:1984bagmodel,Hajduk:1984Skyrme,Schutz:1998Dynamical,Krehl:2000Dynamical,Matsuyama:2007Dynamical,Kamano:2010Dynamical,Kamano:2013Dynamical,Barnes:1983QQG,Golowich:1983QQG,Kisslinger:1995QQG,Roberts:2013RoperLQCD,Kiratidis:2015RoperLQCD}. A prevailing view of the Roper resonance is that it is the first radial excitation of the nucleon~\cite{Burkert:2017djo}. In refined constituent quark models, the Roper is described as a core composed of three dressed quarks, which is then augmented by a surrounding meson cloud. The interaction with the meson cloud modifies the Roper’s core mass, reducing it from approximately 1.75 GeV to its physical mass of 1.44 GeV, thereby resolving the long-standing Roper mass problem.

A significant contribution to this issue came from an analysis using a dynamical coupled-channels model, constrained by experimental scattering data, conducted by Suzuki \textit{et al.}~\cite{Suzuki:2010PoleTrajectory}. Their study revealed that the resonance poles — specifically, the two associated with the Roper — originate from a bare excited nucleon state with a mass of 1.763 GeV. This conclusion was reached by tracing the origins of the resonance through pole trajectories in the complex-energy plane, obtained by varying the interaction strengths from 0\% to 100\%.

With lattice QCD, a more comprehensive understanding of resonances is now possible by incorporating first-principles calculations into the analysis. Lattice studies have found that the first radially excited nucleon state lies at approximately 2  GeV~\cite{Mahbub:2012RoperLQCD,Edwards:2011RoperLattice,Engel:2013RoperLattice,Alexandrou:2015RoperLattice,Liu:2017Roper,Khan:2021RoperLattice}, which is significantly higher than the supposed Roper core mass of 1.75 GeV, and even further from the physical Roper mass of 1.44 GeV.

Lüscher’s formalism allows for a direct connection between finite-volume lattice spectra and infinite-volume scattering observables. In our analysis, we use Hamiltonian Effective Field Theory (HEFT), which has been shown to produce an equivalent relationship to Lüscher’s formalism, to study the Roper resonance as well as the subsequent positive-parity nucleon resonances, $N(1710)$ and/or $N(1880)$.

Here, we provide the first comprehensive study of the positive parity nucleon spectrum, where our Hamiltonian formulation includes the relevant interactions between the ground state nucleon, the first radially excited nucleon state, and various coupled-channel states. Our model is constrained by both experimental scattering data and lattice QCD results.

The structure of this paper is as follows. In Sec.~\ref{sec:theoretical_framework}, we introduce the theoretical framework of the HEFT formalism for a general number of bare states and two-particle scattering (coupled) channels. We outline how HEFT model parameters are defined and used to calculate experimental phase shifts, inelasticities, and resonance positions. Additionally, we present several criteria for constraining the HEFT parameters, ensuring that the calculations adequately reproduce both experimental scattering observables and lattice QCD results. Finally, we summarize the model dependencies (or independencies) of these calculations and evaluate whether they lead to meaningful physical interpretations.

In Sec.~\ref{sec:analysis_of_the_roper}, we present both infinite- and finite-volume results of HEFT using two parameter sets derived from slightly different origins but constrained by the same experimental data and lattice results. We demonstrate that these results successfully reproduce the constraining data, with only mild model dependence manifesting in the finite-volume results. We then discuss the physical interpretation of the Roper resonance and the next two excited nucleon states, the $N(1710)$ and $N(1880)$. The section concludes with an analysis of resonance origins through pole tracing, as previously explored in Ref.~\cite{Suzuki:2010PoleTrajectory}.

In Sec.~\ref{section:conclusion}, we summarize our key findings and present our concluding remarks.

% ---------------------------------------------------------------------------- %
% --------------------------- Theoretical framework -------------------------- %
% ---------------------------------------------------------------------------- %
\section{Theoretical framework}\label{sec:theoretical_framework}

\subsection{Hamiltonian Effective Field Theory}

The Hamiltonian describing an excited baryon embedded in a set of coupled meson-baryon channels can be written as the sum of a non-interacting component, $H_{ 0 }$, and an interaction term, $H_{ I }$,
\begin{equation}
    H = H_{ 0 } + H_{ I } \, .
\end{equation}
The first contribution to $H_0$ describes one or more bare single-particle baryon states, $\ket{ B_{ 0 } }$, which one may interpret as a quark model like baryon state. The second contribution involves the non-interacting coupled two-particle states (denoted by Greek indices) $\alpha$, $\ket{ \alpha( \bm{ k } ) }$, consisting of a baryon $B$ and a meson $M$. Of course, one may also have contributions from states with three or more particles. For the present these are represented by resonant two-body channels involving a baryon resonance, such as the $\Delta$, or meson resonance, such as the $\sigma$ meson.  

While one can work in a moving frame~\cite{Li:2021HEFTmoving}, here we choose the center of mass frame with back-to-back momenta. Then the non-interacting Hamiltonian in HEFT is given by
\begin{align*}\label{eq:free_Hamiltonian}
    H_{ 0 } &= 
    \sum_{ B_{ 0 } } \ket{ B_{ 0 } }\, m_{ B_{ 0 } }\, \bra{ B_{ 0 } } \\
    &+ \sum_{ \alpha } \int d^{ 3 } k\, \ket{ \alpha( \bm{ k } ) }\, \omega_{ \alpha }( \bm{ k } )\, \bra{ \alpha( \bm{ k } ) } \, , 
    \numberthis
\end{align*}
where the subscript 0 denotes a bare baryon state, $m_{ H }$ is the mass of hadron $H$, and $\omega_{ \alpha }( \bm{ k } )$ is the relativistic energy of the two-particle state in channel $\alpha$, given by $\omega_{ \alpha }( \bm{ k } ) = \sqrt{ \bm{ k }^{ 2 } + m_{ B_{ \alpha } }^{ 2 } } + \sqrt{ \bm{ k }^{ 2 } + m_{ M_{ \alpha } }^{ 2 } }$. In this study, we choose the normalization convention $\left\langle B_{ 0 } | B_{ 0 }' \right\rangle = \delta_{ B_{ 0 }, B_{ 0 }' }$, $\left\langle \alpha( \bm{ k } ) | \beta( \bm{ k }' ) \right\rangle = \delta_{ \alpha, \beta }\, \delta( \bm{ k } - \bm{ k }' )$, and $\left\langle B_{ 0 } | \alpha( \bm{ k } )\right\rangle = 0$.

The interaction Hamiltonian consists of two energy-independent terms
\begin{equation}\label{eq:interaction_Hamiltonian}
    H_{ I } = g + v \, .
\end{equation}
The interaction $g$, known as the 1-to-2 interaction, describes the interaction between a bare baryon state and a two-particle state,
\begin{equation}
    g = \sum_{ \alpha, B_{ 0 } } \int d^{ 3 } k 
    \left( \, \ket{ B_{ 0 } }\, G_{ \alpha }^{ B_{ 0 } }( \bm{ k } )\, \bra{ \alpha( \bm{ k } ) } + \text{h.c.}\, \right) \, ,
\end{equation}
where $G_{ \alpha }^{ B_{ 0 } }( \bm{ k } )$ is the momentum dependent vertex function of the 1-to-2 interaction, and h.c. denotes the Hermitian conjugate of the first term. 

The interaction $v$, referred to as the 2-to-2 interaction, describes the interaction between two two-particle states,
\begin{equation}
    v = \sum_{ \alpha, \beta } \int d^{ 3 } k  \int d^{ 3 } k'\, \ket{ \alpha( \bm{ k } ) }\, V_{ \alpha \beta }( \bm{ k }, \bm{ k }' )\, \bra{ \beta( \bm{ k }' ) } \, ,
\end{equation}
where $V_{ \alpha \beta }( \bm{ k }, \bm{ k }' )$ is the momentum-dependent vertex function of the 2-to-2 interaction. 

The forms of the 1-to-2 vertex functions match the vertex interaction of chiral effective field theory ($\chi$EFT). Specifically, for the interaction between the bare baryon state $B_{ 0 }$ and the two-particle channel $\alpha$, the vertex function takes the form
\begin{equation}\label{eq:infinite_volume_1_to_2_vertex_function}
    G_{ \alpha }^{ B_{ 0 } }( \bm{ k } ) =
    \frac{ g_{ \alpha }^{ B_{ 0 } } }{ 2 \pi } \left( \frac{ k }{ f_{ \pi } } \right)^{ l_{ \alpha } }
    \frac{ 1 }{ \sqrt{ \omega_{ M_{ \alpha } }( \bm{ k } ) } } \, .
\end{equation}
Here, $f_{ \pi } = 92.4$ MeV is the pion decay constant, $g_{ \alpha }^{ B_{ 0 } }$ is the dimensionless coupling constant, $l_{ \alpha }$ is the orbital angular momentum of the coupled channel, and $\omega_{ M_{ \alpha } }$ is the energy of the meson $\omega_{ M_{ \alpha } }( \bm{ k } ) \equiv \sqrt{ \bm{ k }^{ 2 } + m_{ M_{ \alpha } }^{ 2 } }$. 

Integrals involving these vertex functions are generally ultraviolet divergent. Thus, following previous HEFT analyses, we use the finite range regulator (FRR) to regulate divergent integrals~\cite{Leinweber:1999ig,Donoghue:1998bs}. Each vertex function, $G_{ \alpha }^{ B_{ 0 } }( \bm{ k } )$, has a corresponding regulator, which for $l_\alpha \, = \, 0$ or $1$ is taken to be a dipole
\begin{equation}\label{eq:1_to_2_regulator}
    u_{ \alpha }^{ B_{ 0 } }( k ) = 
    \left[ 1 + \left( \frac{ k }{ \Lambda_{ \alpha }^{ \! B_{ 0 } } } \right)^{ \! 2 }\ \right]^{ -2 } \, .
\end{equation}
Here $\Lambda_{ \alpha }^{ \! B_0 }$ is the regulator cutoff mass. These phenomenological regulator cutoffs, along with the bare baryon masses and dimensionless coupling constants, are free parameters.

The 2-to-2 interactions are defined in an analogy to the 1-to-2 interaction defined in Eq.~\eqref{eq:infinite_volume_1_to_2_vertex_function}. For interactions between two-particle states $\alpha$ and $\beta$, the 2-to-2 vertex functions are given by
\begin{equation}\label{eq:infinite_volume_2_to_2_vertex_function}
    V_{ \alpha \beta }( \bm{ k }, \bm{ k }' ) = \frac{ v_{ \alpha \beta } }{ 4 \pi^{ 2 } }
    \left( \frac{ k }{ f_{ \pi } } \right)^{ l_{ \alpha } } \left( \frac{ k }{ f_{ \pi } } \right)^{ l_{ \beta } }
    \frac{ 1 }{ \omega_{ M_{ \alpha } }( \bm{ k } ) } \frac{ 1 }{ \omega_{ M_{ \beta } }( \bm{ k }' ) } \, ,
\end{equation}
where $v_{ \alpha \beta }$ is the dimensionless coupling constant. These vertex functions also have corresponding dipole regulators $u_{ \alpha }( k )$, with a regulator cutoff mass, $\Lambda_\alpha$, specific to the two-particle channel. With these definitions, practical calculations of integrals involve the replacements of the vertex functions by
\begin{align*}
    G_{ \alpha }^{ B_{ 0 } }( \bm{ k } ) 
    &\longrightarrow G_{ \alpha }^{ B_{ 0 } }( \bm{ k } )\, u_{ \alpha }^{ B_{ 0 } }( k ), \\
    V_{ \alpha \beta }( \bm{ k }, \bm{ k }' ) 
    &\longrightarrow V_{ \alpha \beta }( \bm{ k }, \bm{ k }' )\, u_{ \alpha }( k )\, u_{ \beta }( k' ) \, . 
    \numberthis
\end{align*}
%

% ---------------------------------------------------------------------------- %
% -------------- HEFT Parameter Constraints and Model Dependence ------------- %
% ---------------------------------------------------------------------------- %
\subsection{HEFT Parameter Constraints and Model Dependence}

Calculations and analyses in HEFT require the determination of the free parameters in the Hamiltonian. These parameters are tuned to suitably reproduce experimental scattering and lattice QCD results. In particular, the parameters are constrained so as to reproduce the following results:
\begin{enumerate}
    \item Experimental scattering observables.
    \item Pole positions of bound and resonance states.
    \item Eigenstates of lattice QCD.
\end{enumerate}
To impose these constraints within HEFT, calculations are performed in both infinite and finite volume. Specifically, the first two constraints are addressed in infinite-volume, while the third is evaluated in finite-volume. 

There have been a number of studies examining the potential model independence of HEFT~\cite{Wu:2014HEFTtoLuscher, Abell:2022Regularization, Hockley:2024DeltaBaryon}. Here, we summarize the key findings.
HEFT demonstrates a correspondence with Lüscher's formalism, as noted in Refs.~\cite{Wu:2014HEFTtoLuscher, Hall:2013DeltaToPiN}. Through this connection, HEFT establishes a model-independent relationship between experimental scattering observables and the finite-volume spectrum at physical masses.
By constraining the HEFT parameters using experimental results, the framework can make accurate predictions — subject to the quality of the experimental data — for the finite-volume spectrum across different lattice volumes and quark masses.

One may also hope to obtain some insight into the nature of baryon resonances by examining the eigenvectors of the finite volume Hamiltonian. Naturally, there is a degree of model dependence in the extraction of the basis state compositions for finite-volume energy eigenstates. Despite this, it has been shown that the compositions generally align with those derived from lattice correlation matrices~\cite{Abell:2023nex}, which are, in turn, dependent on the interpolating fields considered. The resulting decompositions do indeed appear to offer valuable insights into hadron structure.

Finally, depending on the number of free HEFT parameters and the extent to which the experimental results and lattice QCD calculations are reproduced, multiple valid solutions may exist. However, for the cases considered so far, the three constraints appear sufficient to provide a unique and consistent physical interpretation of the nature of the resonances across these solutions.

% ---------------------------------------------------------------------------- %
% ----------------------- Infinite volume calculations ----------------------- %
% ---------------------------------------------------------------------------- %
\subsection{Infinite-volume calculations}

Calculations of scattering variables and resonance extractions in HEFT are performed via the computation of a scattering $T$-matrix. From the interaction Hamiltonian of Eq.~\eqref{eq:interaction_Hamiltonian}, we define an energy dependent 2-to-2 effective potential as
\begin{equation}\label{eq:effective_potential}
    \tilde{ V }_{ \alpha \beta }( \bm{ k }, \bm{ k }'; E ) = \sum_{ B_{ 0 } } 
    \frac{ G_{ \, \alpha }^{ B_{ 0 }\, \dagger }( \bm{ k } )\, G_{ \beta }^{ B_{ 0 } }( \bm{ k } ) }{ E - m_{ B_{ 0 } } }
    + V_{ \alpha \beta }( \bm{ k }, \bm{ k }' ) \, .
\end{equation}
Using this effective potential, we write an integral equation for the $T$-matrix (for each partial wave, suppressing angular momentum indices)
\begin{align*}\label{eq:T_matrix}
    T_{ \alpha \beta }( k, k'; E ) &= \tilde{ V }_{ \alpha \beta }( k, k'; E ) \\
    &+ \sum_{ \gamma } \int d q\, q^{ 2 }\,
    \frac{ \tilde{ V }_{ \alpha \gamma }( k, q; E )\, T_{ \gamma \beta }( q, k'; E ) }
    { E - \omega_{ \gamma }( q ) + i \varepsilon }, \numberthis
\end{align*}
imposing spherically symmetry for the effective potential. The $S$-matrix bears a trivial relationship with the $T$-matrix, given by
\begin{equation}
    S_{ \alpha \beta }( E ) = \delta_{ \alpha \beta }
    - i\, 2 \pi \sqrt{ \rho_{ \alpha }\, \rho_{ \beta } }\ 
    T_{ \alpha \beta }( k_{ \text{on}, \alpha }, k_{ \text{on}, \beta }\,; E ) \, ,
\end{equation}
where $k_{ \text{on}, \gamma }$ is the back-to-back on-shell momentum carried by each hadron in channel $\gamma$, and $\rho_{ \gamma } \equiv \rho_{ \gamma }( k_{ \text{on}, \gamma }\, ; E )$ is the density of states
\begin{equation}
    \rho_{ \gamma } = 
    \frac{ \sqrt{ k_{ \text{on}, \gamma }^{ 2 } + m_{ B_{ \gamma } }^{ 2 }  }\, \sqrt{ k_{ \text{on}, \gamma }^{ 2 } + m_{ M_{ \gamma } }^{ 2 } } }{ E }\,
    k_{ \text{on}, \gamma } \, .
\end{equation}
The scattering phase shifts, $\delta$, and inelasticities, $\eta$, are computed from the diagonal elements of the $S$-matrix, where
\begin{equation}
    S_{ \alpha \alpha }( E ) = \eta_{ \alpha } \exp\left( \,i\, 2 \delta_{ \alpha }\, \right) \, .
\end{equation}

As bound and resonant states manifest as poles of the $T$-matrix, we need to consider how to locate those poles.
Noting the definition of the effective potential in Eq.~\eqref{eq:effective_potential} and the integral equation for the $T$-matrix in Eq.~\eqref{eq:T_matrix}, the $T$-matrix can be expressed as a sum of two $T$-matrices: one involving the 1-to-2 interaction, $T_{ \alpha \beta }^{ \text{bare} }$, and the other involving only the 2-to-2 interaction, $t_{ \alpha \beta }$. Explicitly we find
\begin{equation}\label{eq:T_matrix_sum}
    T_{ \alpha \beta }( k, k'; E ) = 
    t_{ \alpha \beta }( k, k'; E ) + T_{ \alpha \beta }^{ \text{bare} }( k, k'; E ) \, .
\end{equation}
From the above, a pole in the 2-to-2 $T$-matrix, $t_{ \alpha \beta }$, should correspond to a pole in the full $T$-matrix. However, in HEFT, the 2-to-2 $T$-matrix in the sum is ``canceled" by a corresponding term in the bare $T$-matrix. As a result, pure 2-to-2 poles do not exist in the full system. This does not imply, however, that information about the 2-to-2 poles is lost. Instead, it is modified through the effective 1-to-2 interactions encoded in the bare $T$-matrix. The bare $T$-matrix is given by
\begin{equation}
    T_{ \alpha \beta }^{ \text{bare} }( k, k'; E ) = \sum_{ B_{ 0 }, B_{ 0 }' }
    \mathcal{ G }_{ \, \alpha }^{ B_{ 0 } }( k; E )\, A_{ B_{ 0 }, B_{ 0 }' }( E )\, \mathcal{ G }_{ \, \beta }^{ B_{ 0 }' }( k'; E ) \, ,
\end{equation}
where $A$ is the dressed propagator, and the effective 1-to-2 interaction $\mathcal{ G }_{ \, \alpha }^{ B_{ 0 } }$ is defined as
\begin{equation}
    \mathcal{ G }_{ \, \alpha }^{ B_{ 0 } }( k; E ) \equiv G_{ \, \alpha }^{ B_{ 0 } }( k ) 
    + \sum_{ \gamma } \int_{ C_{ \gamma } } dq \, q^{ 2 }\,
    \frac{ t_{ \alpha \gamma }( k, q; E )\, G_{ \gamma }^{ B_{ 0 } }( q ) }{ E - \omega_{ \gamma } + i \varepsilon } \, ,
\end{equation}
where $C_{ \gamma }$ represents the contour of integration. 

The dependence of $t_{ \alpha \beta }$ on $\mathcal{ G }_{ \, \alpha }^{ B_{ 0 } }$ highlights how information about the 2-to-2 poles is retained in the full system. In fact, in the limit where the 1-to-2 coupling constants become small, the full system effectively reproduces the pure 2-to-2 poles.

The poles of the full $T$-matrix correspond exclusively to the unique poles in the bare $T$-matrix. These unique poles in the bare $T$-matrix are located within the dressed propagator, and they can be found by solving for the complex energy, $E$, that satisfies $\det \! \left( \, A^{ -1 }\, \right) = 0$, where $A^{ -1 }$ is given by
\begin{equation}
    A_{ B_{ 0 }, B_{ 0 }' }^{ -1 }( E ) = 
    \delta_{ B_{ 0 }, B_{ 0 }' }\left( E - m_{ B_{ 0 } } \right) - \tilde{ \Sigma }_{ B_{ 0 }, B_{ 0 }' }( E ) \, .
\end{equation}
Here, $\tilde{ \Sigma }( E )$ is the total self-energy contribution arising from all possible 1-to-2 and 2-to-2 one loop diagrams. The self-energies contain integrals (or products of integrals),  $\mathcal{ I }$, which can be generically written as
\begin{equation}\label{eq:generic_self_energy_integral}
    \mathcal{ I }( E ) = \sum_{ \gamma } \int_{ C_{ \gamma } } dq \, q^{ 2 }\, 
    \frac{ f_{ \gamma }( q )\, g_{ \gamma }( q ) }{ E - \omega_{ \gamma } + i \varepsilon } \, ,
\end{equation}
where, unique to each channel $\gamma$, 
$f_{ \gamma }$ and $g_{ \gamma }$ are functions of complex momentum, $q$. The choice of $C_{ \gamma }$ directly influences the value of $\mathcal{ I }( E )$, and consequently, the location of the poles.

% ---------------------------------------------------------------------------- %
% ------------------------ Finite volume calculations ------------------------ %
% ---------------------------------------------------------------------------- %
\subsection{Finite-volume calculations}

The quantization of momenta in a cubic lattice of volume $L^{ 3 }$, is given by
\begin{equation}
    \bm{ k }_{ \bm{ n } } = \frac{ 2 \pi }{ L }\, \bm{ n } \, ,
\end{equation}
where $L$ is the lattice extent and $\bm{ n } \in \mathbb{ Z }^{ 3 }$. Thus, integrals over momenta in $\mathbb{ R }^{ 3 }$ are replaced by sums over the quantized momenta in $\mathbb{ Z }^{ 3 }$. Analogous to the spherical symmetry in the infinite-volume, we impose cubic symmetry on the finite-volume system. With these symmetries, a three-dimensional integral or sum reduces to a one-dimensional one. We denote the discretized momentum by a single index $n$, where $k_{ n } = \left| \bm{ k_{ \bm{ n } } } \right|$. This allows us to establish the connection
\begin{equation}\label{eq:infinite_volume_to_finite_volume}
    \int_{ \mathbb{ R }^{ 3 } } d^{ 3 } k = 4 \pi \int_{ 0 }^{ \infty } d k\,  k^{ 2 } 
    \longrightarrow 
    \left( \frac{ 2 \pi }{ L } \right)^{ 3 } \sum_{ n \in \mathbb{ S } } r_{ 3 }( n ) \, ,
\end{equation}
where $n = \left| \bm{ n } \right|$, and $r_{ j }(n)$ (or $C_{ j }( n )$) is the sum of squares function, which counts the number of ways natural number $n$ can be expressed as a sum of $j$ squared integers. Here, $\mathbb{ S } = \mathbb{ Z }_{ 0 }^{ + }$ for $S$-wave two-particle channels, and $\mathbb{ S } = \mathbb{ Z }^{ + }$ for non-$S$-wave channels.

With discretization of spacetime and a finite lattice extent, momentum dependent particle states and operators can be represented by finite dimensional complex-valued vectors and matrices, respectively. To ensure the Hamiltonian matrix is finite in size, a maximum momentum cutoff $k_{ n_{ \text{max} } }$ is required. This cutoff defines the maximum number of basis states and thus determines the size of the Hamiltonian (matrix). For practical calculations, $k_{ n_{ \text{max} } }$ is chosen to satisfy two criteria: firstly, variations in $k_{ n_{ \text{max} } }$ do not appreciably alter the solutions of the Hamiltonian, and secondly, the size of the Hamiltonian should be minimized to reduce computing time. The momentum cutoff is governed by the regulator $u$ and its regulator cutoff mass, $\Lambda$. A suitable value of $k_{ n_{ \text{max} } }$ can be obtained by solving $u( k_{ n_{ \text{max} } }, \Lambda ) = u_{ \text{min} }$, where $u_{ \text{min} }$ satisfies the aforementioned criteria. For the dipole regulator, these relationships are given by
\begin{alignat*}{2}
    & k_{ n_{ \text{max} } } &&= \Lambda\, \sqrt{ u_{ \text{min} }^{ -1 / 2 } - 1 }\, , \\
    & n_{ \text{max} }       &&= \left( \frac{ k_{ n_{ \text{max} } }\, L }{ 2 \pi } \right)^{ 2 } \, . \numberthis
\end{alignat*}
Reference~\cite{Abell:2022Regularization} investigated the finite-range regularization of HEFT for the $\Delta$ resonance. In Fig.~2 of that reference, the finite-volume HEFT energy eigenstates are shown as a function of $u_{\text{min}}$ for a dipole regulator. It is found that a value of $u_{ \text{min} } = 0.01$ adequately satisfies the two criteria, and we adopt this value in our analysis.

The extraction of bound and resonance state energies in finite-volume HEFT is performed by solving the eigenvalue equation of the Hamiltonian
\begin{equation}\label{eq:eigenvalue_equation}
    \det \! \left( H - E_{ i }\, \mathds{ 1 } \right) = 0 \, .
\end{equation}
The energy spectrum produced should be in accord with the bound and resonance state masses of $\chi$EFT in finite-volume. For this to be achieved, one needs to appropriately scale the vertex functions of HEFT. With the replacement of Eq.~\eqref{eq:infinite_volume_to_finite_volume}, the vertex functions in a finite-volume are
\begin{align*}
    \bar{ G }_{ \alpha }^{ B_{ 0 } }( k_{ n } ) 
    &= \sqrt{ \frac{ r_{ 3 }( n ) }{ 4 \pi } } \left( \frac{ 2 \pi }{ L } \right)^{ \! 3/2 } 
    G_{ \alpha }^{ B_{ 0 } }( k_{ n } ), \\
    \bar{ V }_{ \alpha \beta }( k_{ m }, k_{ n } )
    &=  \frac{ \sqrt{ r_{ 3 }( m )\, r_{ 3 }( n ) } }{ 4 \pi }
    \left( \frac{ 2 \pi }{ L } \right)^{ \! 3 } V_{ \alpha \beta }( k_{ m }, k_{ n } ) \, , \numberthis
\end{align*}
with the infinite-volume vertex functions provided in Eqs.~\eqref{eq:infinite_volume_1_to_2_vertex_function} and \eqref{eq:infinite_volume_2_to_2_vertex_function} and including the corresponding regulators.

Finally, to impose more of the lattice QCD constraints on the HEFT parameters, it is necessary to extend the formalism to unphysical hadron masses. This is done by expressing the hadron masses in the two-body states as a (truncated) power series in pion mass squared (or light quark mass), motivated by $\chi$EFT. Explicitly, for some physical or bare hadron $H$, its mass can be written as
\begin{equation}\label{eq:hadron_mass_expansion}
    m_{ H }\left( m_{ \pi }^{ 2 } \right) = 
    \left. m_{ H } \right|_{ \text{e.p.} } 
    + \sum_{ p = 1 }^{ p_{ \text{max} } } \alpha_{ H, p } 
    \left[ \, \left( m_{ \pi }^{ 2 } \right)^{ p } - \left. \left( m_{ \pi }^{ 2 } \right)^{ p } \right|_{ \text{e.p.} }\, \right] \, .
\end{equation}
 Here, $\alpha_{ H, p }$ denotes the fit slope coefficients, and e.p. stands for the pion mass evaluation point. For example, if $\text{e.p.} = 0.2$ GeV and $\left. m_{ H } \right|_{ \text{e.p.} } = 1.0$ GeV, this means that the hadron mass is $1.0$ GeV at $m_{ \pi } = 0.2$ GeV. Typically $p_{ \text{max} }$ does not exceed 2, so that the bare baryon mass changes either linearly or quadratically with quark mass or $m_{ \pi }^{ 2 }$. The evaluation point for different hadrons may, but need not necessarily, be the same, depending on the availability of the lattice QCD results. It should be noted that we allow for finite-volume effects, such that physical hadron masses and the finite-volume $m_{ H }$ extrapolated at the physical point are generally not the same.

% ---------------------------------------------------------------------------- %
% --------------------------- Eigenvector analysis --------------------------- %
% ---------------------------------------------------------------------------- %
\subsection{Eigenvector analysis}

The HEFT formalism provides a clear and effective framework for understanding bound and resonance states in finite-volume. As discussed in the previous section, these states in finite-volume are extracted by solving the eigenvalue equation given in Eq.~\eqref{eq:eigenvalue_equation}. In terms of the basis kets, this equation takes the form $H \ket{ E_{ i } } = E_{ i } \ket{ E_{ i } }$, where $E_{ i }$ is the energy eigenvalue of the $i$-th eigenstate $\ket{ E_{ i } }$.

The vector space is spanned by the bare baryon and two-particle basis kets, which satisfy normalization conditions analogous to those in infinite-volume. These conditions are $\left\langle B_{ 0 } | B_{ 0 }' \right\rangle = \delta_{ B_{ 0 }, B_{ 0 }' }$, $\left\langle \alpha( k_{ m } ) | \beta( k_{ n } ) \right\rangle = \delta_{ \alpha, \beta }\, \delta_{ m, n }$, and $\left\langle B_{ 0 } | \alpha( k_{ n } )\right\rangle = 0$. Therefore, any state within this vector space can be written as a linear combination of these basis states. Accordingly, the energy eigenket $\ket{ E_{ i } }$ can be written as
\begin{equation}\label{eq:energy_eigenket}
    \ket{ E_{ i } } = 
    \sum_{ j = 1 }^{ n_{ b } } b_{ \, i }^{ ( j ) } \ket{ B_{ 0 }^{ ( j ) } } +
    \sum_{ j = 1 }^{ n_{ c } } \sum_{ n = 1 }^{ n_{ \text{max} } } c_{ \, i, n }^{ ( j ) } \ket{ \alpha^{ ( j ) }( k_{ n } ) } \, ,
\end{equation}
where the system contains $n_{ b }$ bare baryon states and $n_{ c }$ two-particle states, and the coefficients $b_{ \, i }^{ ( j ) }$ and $c_{ \, i, n }^{ ( j ) }$ are defined by $b_{ \, i }^{ ( j ) } \equiv \langle B_{ 0 }^{ ( j ) } | E_{ i } \rangle$ and $c_{ \, i, n }^{ ( j ) } \equiv \left\langle \alpha^{ ( j ) }( k_{ n } ) \middle| 
E_{ i } \right\rangle$, respectively. Given this formalism, the contribution of each basis state to a given energy level is directly given by the squared modulus of the corresponding coefficient. For instance, to compute the contribution of the first bare baryon state $\ket{ B_{ 0 } }$ to the first energy eigenstate $\ket{ E_{ 1 } }$, we calculate the (squared) overlap $\left| \left\langle B_{ 0 } \middle| E_{ 1 } \right\rangle \right|^{ 2 } = | b_{ \, 1 }^{ ( 1 ) } |^{ 2 }$.

% ---------------------------------------------------------------------------- %
% -------------------------- HEFT Hamiltonian matrix ------------------------- %
% ---------------------------------------------------------------------------- %
\subsection{Hamiltonian matrix}

For a system with $n_{ b }$ bare states and $n_{ c }$ two-particle states, the finite-volume 
non-interacting Hamiltonian matrix can be written as
\begin{align*}
    \mathbf{ H }_{ 0 } = 
    \text{diag}
    & \left[ m_{ B_{ 0 }^{ ( 1 ) } }, \cdots, m_{ B_{ 0 }^{ ( n_{ b } ) } }, \omega_{ \alpha^{ ( 1 ) } }( k_{ 0 } ), \cdots, \omega_{ \alpha^{ ( n_{ c } ) } }( k_{ 0 } ), \right. \\
    & \mkern15mu \left. \cdots, \omega_{ \alpha^{ ( 1 ) } }( k_{ n_{ \text{max} } } ), \cdots, \omega_{ \alpha^{ ( n_{ c } ) } }( k_{ n_{ \text{max} } } ) \vphantom{ m_{ B_{ 0 }^{ ( n_{ b } ) } } } \right].
\end{align*}
Similarly, for the interaction Hamiltonian, we write
\begin{equation}
    \mathbf{ H }_{ I } =
    \begin{pNiceArray}{c|ccc}
        \bm{ 0 }_{ n_{ b } \times n_{ b } } & \bar{ \mathbf{ G } }( k_{ 0 } ) & \cdots & \bar{ \mathbf{ G } }( k_{ \text{max} } ) \\[4pt] \hline \\[-7pt]
        \bar{ \mathbf{ G } }^{ T }( k_{ 0 } ) & \bar{ \mathbf{ V } }( k_{ 0 }, k_{ 0 } ) & \cdots & \bar{ \mathbf{ V } }( k_{ 0 }, k_{ \text{max} } ) \\
        \vdots & \vdots & \ddots & \vdots \\
        \bar{ \mathbf{ G } }^{ T }( k_{ \text{max} } ) & \bar{ \mathbf{ V } }( k_{ \text{max} }, k_{ 0 } ) & \cdots & \bar{ \mathbf{ V } }( k_{ \text{max} }, k_{ \text{max} } )
    \end{pNiceArray},
\end{equation}
where $k_{ \text{max} } \equiv k_{ n_{ \text{max} } }$. The superscript $T$ denotes the matrix transpose, and the matrices $\bar{ \mathbf{ G } }$ and $\bar{ \mathbf{ V } }$ are defined as follows
\begin{equation}
    \bar{ \mathbf{ G } }( k_{ n } ) \equiv
    \begin{pmatrix}
        \bar{ G }_{ 1 }^{ 1 }( k_{ n } )       & \cdots & \bar{ G }_{ n_{ c } }^{ 1 }( k_{ n } ) \\
        \vdots                                 & \ddots & \vdots \\
        \bar{ G }_{ 1 }^{ n_{ b } }( k_{ n } ) & \cdots & \bar{ G }_{ n_{ c } }^{ n_{ b } }( k_{ n } )       
    \end{pmatrix} \, ,
\end{equation}
and
\begin{equation}
    \bar{ \mathbf{ V } }( k_{ m }, k_{ n } ) \equiv
    \begin{pmatrix}
        \bar{ V }_{ 1, 1 }( k_{ m }, k_{ n } )       & \cdots & \bar{ V }_{ 1, n_{ c } }( k_{ m }, k_{ n } ) \\
        \vdots                                       & \ddots & \vdots \\
        \bar{ V }_{ n_{ c }, 1 }( k_{ m }, k_{ n } ) & \cdots & \bar{ V }_{ n_{ c }, n_{ c } }( k_{ m }, k_{ n } )      
    \end{pmatrix} \, .
\end{equation}
For brevity, the subscripts and superscripts of $\bar{ G }_{ i }^{ j }$ denote the $i$-th two-particle channel and $j$-th bare baryon state (as in $\bar{ G }_{ \alpha }^{ B_{ 0 } }$), respectively. For $\bar{ V }$, the two subscripts specify the two two-particle states (as in $\bar{ V }_{ \alpha \beta }$). We reiterate that the value of $k_{ 0 }$ varies depending on the partial wave of the two-particle channel, specifically,
\begin{equation}
    k_{ 0 } =
    \begin{cases}
        0                   & \text{if } S \text{-wave}, \\
        \frac{ 2 \pi }{ L } & \text{otherwise} \, .
    \end{cases}
\end{equation}
%

% ---------------------------------------------------------------------------- %
% ----------------------------------- HEFT ----------------------------------- %
% ---------------------------------------------------------------------------- %
\section{Analysis of the Roper resonance}\label{sec:analysis_of_the_roper}

For this study, we consider a system with two bare states, $n_{ b } = 2$, and four two-body channels, $n_{ c } = 4$, abbreviated as 2b4c. The first bare baryon state corresponds to the ground nucleon state $N$, and the second corresponds to the first radially excited state $N^{ * }$, where we label these states $N_{ 1 }$ and $N_{ 2 }$, respectively. For the two-particle states, we consider the channels $\pi N$, $\pi \Delta$, $\eta N$, and $\sigma N$, with the first three channels in $P$-wave and the last in $S$-wave. Using this 2b4c system, we aim to reproduce the ground state nucleon, the Roper resonance $N(1440)$, and the resonance $N(1710)$, and/or $N(1880)$. 

In a 2b4c system, there are at most 32 HEFT parameters that must be determined. In the interest of efficiency and noting that adding additional bare baryon and two-particle states is straightforward in HEFT, we first consider simpler systems, such as 0b4c or 1b4c. These systems involve significantly fewer parameters, and their values can serve as initial estimates for the parameters of the full system.

The experimental scattering phase shifts that we fit are extracted from the partial-wave analyses of $\pi N \to \pi N$ scattering, specifically the WI08 solution for the $P_{ 11 }$ partial wave from Ref.~\cite{GWU:2025ScatteringAmplitudes}. We consider the energy range 1200 MeV to 1900 MeV, which captures the higher energy N(1/2$^{ + }$) resonances.

When constraining the resonance parameters it is important to choose an appropriate integration contour, $C_{ \gamma }$ (see Eq.~\eqref{eq:generic_self_energy_integral}). We focus on poles that affect $\pi N$ scattering observables and are close to the physical region. These poles are found on the second Riemann sheet of the complex energy plane (the unphysical sheet) in the $\pi N$ channel. However, in principle, in the other channels the poles could lie on either the physical sheet, $p$ (the first Riemann sheet), or the unphysical sheet, $u$. To identify unique pole locations, we label the sets of sheets as $\left\lbrace s_{ \pi N }, s_{ \pi \Delta }, s_{ \eta N }, s_{ \sigma N } \right\rbrace$. 

As the Roper resonance lies near the $\pi \Delta$ threshold, two poles are associated with it: one located on the physical and the other on the unphysical $\pi \Delta$ sheet, the $uppu$ and the $uupu$ sheets, respectively. The pole closer to the Roper pole position reported by the Particle Data Group (PDG)~\cite{PDG:2022PoleLocations} is selected as the constraining pole for the HEFT parameters. For the higher energy resonances, $N(1710)$ and/or $N(1880)$, we search for the pole(s) on the $uuuu$-sheet, where all channels are open.

For simplicity, we choose the contour integration path $C_{ \gamma }$, to run along the positive real axis, $q \in [ 0, +\infty )$. To access the poles in the unphysical sheets, the complex momentum is rotated $q \to q\, e^{ i \theta_{ \gamma } }$, where $\theta_{ \gamma }$ is the rotation angle specific to the channel $\gamma$. For most poles in the unphysical sheet, a rotation angle $\theta_{ \gamma } \approx -70^{ \circ }$ is sufficient.

The constraints on HEFT parameters from lattice QCD are obtained by combining results from CSSM~\cite{Liu:2017Roper}, Lang \textit{et al.}~\cite{Lang:2017LatticeResult}, and PACS-CS~\cite{PACSCS:2009LatticeResult}. These parameters are constrained through two key methods utilizing lattice QCD results:
\begin{itemize}
    \item finite volume energy eigenstates are matched with the eigenstates observed in lattice QCD calculations.
    \item HEFT energy eigenstates with significant bare state contributions are aligned with lattice QCD results generated using three-quark interpolators.
\end{itemize}
The second constraint is more nuanced and is achieved by separately tuning both the HEFT parameters and the slope parameters of the bare baryon mass expansion, as given in Eq.~\eqref{eq:hadron_mass_expansion}. 

\begin{table}[t]
    \caption{Table of two-particle channel physical hadron masses $\left. m_{ H } \right|_{ \text{phys} }$ and lattice hadron masses $\left. m_{ H } \right|_{ \text{e.p.} }$, with the corresponding pion mass evaluation point e.p. and slope parameter $\alpha_{ H, p }$, as described in the text below Eq.~\eqref{eq:hadron_mass_expansion}. The variables under lattice heading are only used in the finite-volume calculations, and $\left. m_{ H } \right|_{ \text{phys} }$ is only used in infinite-volume.} 
    \label{tab:two_particle_hadron_masses_and_slopes}
    \begin{ruledtabular}
        \begin{tabular}{ccccc}
        \\[-8pt]
        & {Physical} & \multicolumn{3}{c}{Lattice} \\
        \cmidrule(lr){3-5} \\[-10pt]
        {$H$} & {$\left. m_{ H } \right|_{ \text{phys} }$} & {e.p.} & {$\left. m_{ H } \right|_{ \text{e.p.} }$} & {$\alpha_{ H, 1 }$} \\[2pt]
        & (GeV) & (GeV) & (GeV) & (GeV$^{ -1 }$) \\[2pt]
        \hline \\[-8pt]
        $\eta$   & 0.548 & 0.156 & 0.633 & 0.333 \\
        $\sigma$ & 0.350 & 0.156 & 0.350 & 0.803 \\
        $N$      & 0.939 & 0.156 & 0.977 & 1.209 \\
        $\Delta$ & 1.232 & 0.139 & 1.263 & 0.972
        \end{tabular}
    \end{ruledtabular}
\end{table}

In Table~\ref{tab:two_particle_hadron_masses_and_slopes} we present hadron masses in the two-particle channels in infinite-volume (physical), as well as the mass expansion parameters of the same hadrons in the finite-volume calculations. The values for the nucleon are obtained by  combining results from Refs.~\cite{Lang:2017LatticeResult} and \cite{Liu:2017Roper}, while the values for the $\Delta$ are taken from Ref.~\cite{Hockley:2024DeltaBaryon}. For the $\eta$ and $\sigma$ mesons we use the $\eta$ mass given in Ref.~\cite{PACSCS:2009LatticeResult} and the $\sigma$ mass used in previous HEFT Roper analyses~\cite{Liu:2017Roper, Wu:2017qve}. For the $\eta$ and $\sigma$ slope parameters we use the leading order $\chi$EFT mass expansion, \textit{i.e.} $\alpha_{ \eta, 1 } = 1 / 3$, and $\alpha_{ \sigma, 1 } = 2 / 3\,  \alpha_{ N, 1 }$.

When constraining the HEFT parameters, we construct a combined $\chi^{ 2 }$ function that incorporates both experimental scattering observables and lattice results, but not the bound and resonance pole positions. However, these fits are expected to constrain the HEFT parameters sufficiently to produce the required poles. 

Finding the absolute global minimum of the $\chi^{ 2 }$ function in this 32-dimensional parameter space is challenging. Moreover, without all possible physical constraints, it is not possible to single out a unique solution corresponding to nature. Nevertheless, when appropriately constrained by relevant physical results, it appears that the resulting solutions yield a consistent physical interpretation of the resonances.
\begin{figure*}
    \centering
    \includegraphics[width=\textwidth]{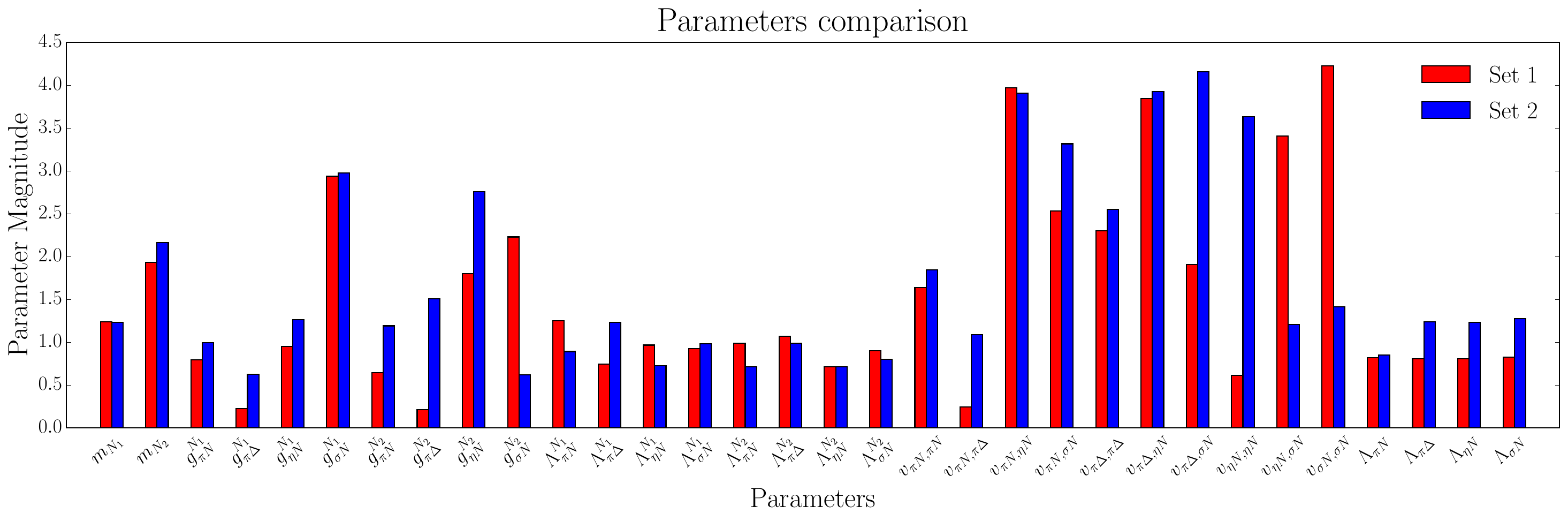}
    \caption{Comparison of the parameter sets 1 and 2. The magnitude of the HEFT parameters are displayed.}
    \label{fig:parameter_bar_plot}
\end{figure*}
%

% ---------------------------------------------------------------------------- %
% ------------------ Infinite-volume fits and pole positions ----------------- %
% ---------------------------------------------------------------------------- %
\subsection{Infinite-volume fits and pole positions}

In this study, we present two parameter sets with distinct origins that reasonably satisfy the constraints on HEFT parameters. These parameter sets are used to explore the degree of model dependence in the physical interpretations of the results.

The first parameter set (Set 1) originates from a 0b4c system, which produces a pole associated with the Roper resonance. This system is then evolved into a 2b4c system by introducing the bare nucleon state $N_{ 1 }$ and the first radially excited state $N_{ 2 }$. In this case it is reasonable to view the Roper as a dynamically generated resonance, as the resonance pole is found in the 2-to-2 $T$-matrix. Of course, this interpretation must be moderated by the realization that within the full 2b4c framework it has a complex state structure incorporating some mixing with the bare baryon states. This is because 2-to-2 poles are modified when embedded in the full system. In essence, the full interaction transforms the resonance into a more intricate and sophisticated structure.

The second parameter set (Set 2) is derived from a 1b4c system, which evolves into a 2b4c system through the introduction of the second bare state, $N_{ 2 }$. In this case, the 2-to-2 couplings differ significantly from those in Set 1, and the Roper resonance emerges only within the dressed propagator $A$. This indicates that, in this scenario, the Roper arises from a more involved interplay of bare states and coupled-channel interactions.

\begin{table}[!t]
    \caption{HEFT parameters for parameter set 1.} 
    \label{tab:parameter_set_1}
    \begin{ruledtabular}
        \begin{tabular}{cccc}
        \multicolumn{2}{c}{1-to-2 $N_{ 1 }$} & \multicolumn{2}{c}{1-to-2 $N_{ 2 }$} \\
        \cmidrule(lr){1-2} \cmidrule(lr){3-4} \\[-10pt]
        Parameter & Value & Parameter & Value \\[2pt]
        \hline \\[-8pt]
        $m_{ N_{ 1 } }$ (GeV)          &  1.237 & $m_{ N_{ 2 } }$ (GeV)          &  1.934 \\[2pt]
        $g_{ \pi N }^{ N_{ 1 } }$      &  0.796 & $g_{ \pi N }^{ N_{ 2 } }$      & -0.646 \\[2pt]
        $g_{ \pi \Delta }^{ N_{ 1 } }$ &  0.223 & $g_{ \pi \Delta }^{ N_{ 2 } }$ & -0.214 \\[2pt]
        $g_{ \eta N }^{ N_{ 1 } }$     & -0.948 & $g_{\eta N }^{ N_{ 2 } }$      &  1.802 \\[2pt]
        $g_{ \sigma N }^{ N_{ 1 } }$   &  2.935 & $g_{ \sigma N }^{ N_{ 2 } }$   &  2.228 \\[2pt]
        $\Lambda_{ \pi N }^{ N_{ 1 } }$ (GeV)      & 1.251 &
        $\Lambda_{ \pi N }^{ N_{ 2 } }$ (GeV)      & 0.989 \\[2pt]
        $\Lambda_{ \pi \Delta }^{ N_{ 1 } }$ (GeV) & 0.745 &
        $\Lambda_{ \pi \Delta }^{ N_{ 2 } }$ (GeV) & 1.067 \\[2pt]
        $\Lambda_{ \eta N }^{ N_{ 1 } }$ (GeV)     & 0.966 &
        $\Lambda_{ \eta N }^{ N_{ 2 } }$ (GeV)     & 0.710 \\[2pt]
        $\Lambda_{ \sigma N }^{ N_{ 1 } }$ (GeV)   & 0.924 &
        $\Lambda_{ \sigma N }^{ N_{ 2 } }$ (GeV)   & 0.900 \\[2pt]
        \hline \\[-8pt]
        & \multicolumn{2}{c}{$\mkern-50mu$ 2-to-2} & \\
        \cmidrule(lr){1-4} \\[-10pt]
        Parameter & Value & Parameter & Value \\[2pt]
        \hline \\[-8pt]
        $v_{ \pi N, \pi N }$           &  1.636 & $v_{ \pi \Delta, \eta N }$   & -3.844 \\
        $v_{ \pi N, \pi \Delta }$      &  0.243 & $v_{ \pi \Delta, \sigma N }$ & -1.907 \\
        $v_{ \pi N, \eta N }$          & -3.969 & $v_{ \eta N, \eta N }$       &  0.613 \\
        $v_{ \pi N, \sigma N }$        &  2.532 & $v_{ \eta N, \sigma N }$     & -3.407 \\
        $v_{ \pi \Delta, \pi \Delta }$ &  2.302 & $v_{ \sigma N, \sigma N }$   &  4.223 \\
        $\Lambda_{ \pi N }$ (GeV)      &  0.821 & $\Lambda_{ \eta N }$ (GeV)   &  0.808 \\
        $\Lambda_{ \pi \Delta }$ (GeV) &  0.807 & $\Lambda_{ \sigma N }$ (GeV) &  0.828 \\
        \end{tabular}
    \end{ruledtabular}
\end{table}
\begin{table}[!t]
    \caption{HEFT parameters for parameter set 2.} 
    \label{tab:parameter_set_2}
    \begin{ruledtabular}
        \begin{tabular}{cccc}
        \multicolumn{2}{c}{1-to-2 $N_{ 1 }$} & \multicolumn{2}{c}{1-to-2 $N_{ 2 }$} \\
        \cmidrule(lr){1-2} \cmidrule(lr){3-4} \\[-10pt]
        Parameter & Value & Parameter & Value \\[2pt]
        \hline \\[-8pt]
        $m_{ N_{ 1 } }$ (GeV)          &  1.234 & $m_{ N_{ 2 } }$ (GeV)          &  2.164 \\[2pt]
        $g_{ \pi N }^{ N_{ 1 } }$      & -0.993 & $g_{ \pi N }^{ N_{ 2 } }$      &  1.191 \\[2pt]
        $g_{ \pi \Delta }^{ N_{ 1 } }$ &  0.625 & $g_{ \pi \Delta }^{ N_{ 2 } }$ & -1.505 \\[2pt]
        $g_{ \eta N }^{ N_{ 1 } }$     & -1.265 & $g_{\eta N }^{ N_{ 2 } }$      &  2.758 \\[2pt]
        $g_{ \sigma N }^{ N_{ 1 } }$   &  2.977 & $g_{ \sigma N }^{ N_{ 2 } }$   &  0.622 \\[2pt]
        $\Lambda_{ \pi N }^{ N_{ 1 } }$ (GeV)      & 0.891 &
        $\Lambda_{ \pi N }^{ N_{ 2 } }$ (GeV)      & 0.711 \\[2pt]
        $\Lambda_{ \pi \Delta }^{ N_{ 1 } }$ (GeV) & 1.232 &
        $\Lambda_{ \pi \Delta }^{ N_{ 2 } }$ (GeV) & 0.986 \\[2pt]
        $\Lambda_{ \eta N }^{ N_{ 1 } }$ (GeV)     & 0.726 &
        $\Lambda_{ \eta N }^{ N_{ 2 } }$ (GeV)     & 0.712 \\[2pt]
        $\Lambda_{ \sigma N }^{ N_{ 1 } }$ (GeV)   & 0.982 &
        $\Lambda_{ \sigma N }^{ N_{ 2 } }$ (GeV)   & 0.801 \\[2pt]
        \hline \\[-8pt]
        & \multicolumn{2}{c}{$\mkern-50mu$ 2-to-2} & \\
        \cmidrule(lr){1-4} \\[-10pt]
        Parameter & Value & Parameter & Value \\[2pt]
        \hline \\[-8pt]
        $v_{ \pi N, \pi N }$           &  1.844 & $v_{ \pi \Delta, \eta N }$   & -3.928 \\
        $v_{ \pi N, \pi \Delta }$      & -1.090 & $v_{ \pi \Delta, \sigma N }$ & -4.155 \\
        $v_{ \pi N, \eta N }$          &  3.904 & $v_{ \eta N, \eta N }$       &  3.633 \\
        $v_{ \pi N, \sigma N }$        &  3.316 & $v_{ \eta N, \sigma N }$     & -1.206 \\
        $v_{ \pi \Delta, \pi \Delta }$ &  2.549 & $v_{ \sigma N, \sigma N }$   &  1.415 \\
        $\Lambda_{ \pi N }$ (GeV)      &  0.850 & $\Lambda_{ \eta N }$ (GeV)   &  1.230 \\
        $\Lambda_{ \pi \Delta }$ (GeV) &  1.238 & $\Lambda_{ \sigma N }$ (GeV) &  1.276 \\
        \end{tabular}
    \end{ruledtabular}
\end{table}

In Tables~\ref{tab:parameter_set_1} and \ref{tab:parameter_set_2}, we present the constrained HEFT parameters for the two parameter sets that we just described. Figure~\ref{fig:parameter_bar_plot} provides a visual comparison of the HEFT parameters for these sets. Here, only the magnitudes of the parameters are shown to highlight the relative contributions of each participating interaction. A brief inspection of the plot reveals that the parameters are generally quite similar, with the exception of about half of the 2-to-2 couplings and certain 1-to-2 couplings. This discrepancy is unsurprising, as the only experimental data used to constrain the HEFT parameters came from $\pi N \to \pi N$ scattering.

\begin{figure*}
    \centering
    \begin{minipage}[h]{.495\textwidth}
        \includegraphics[width=\textwidth]{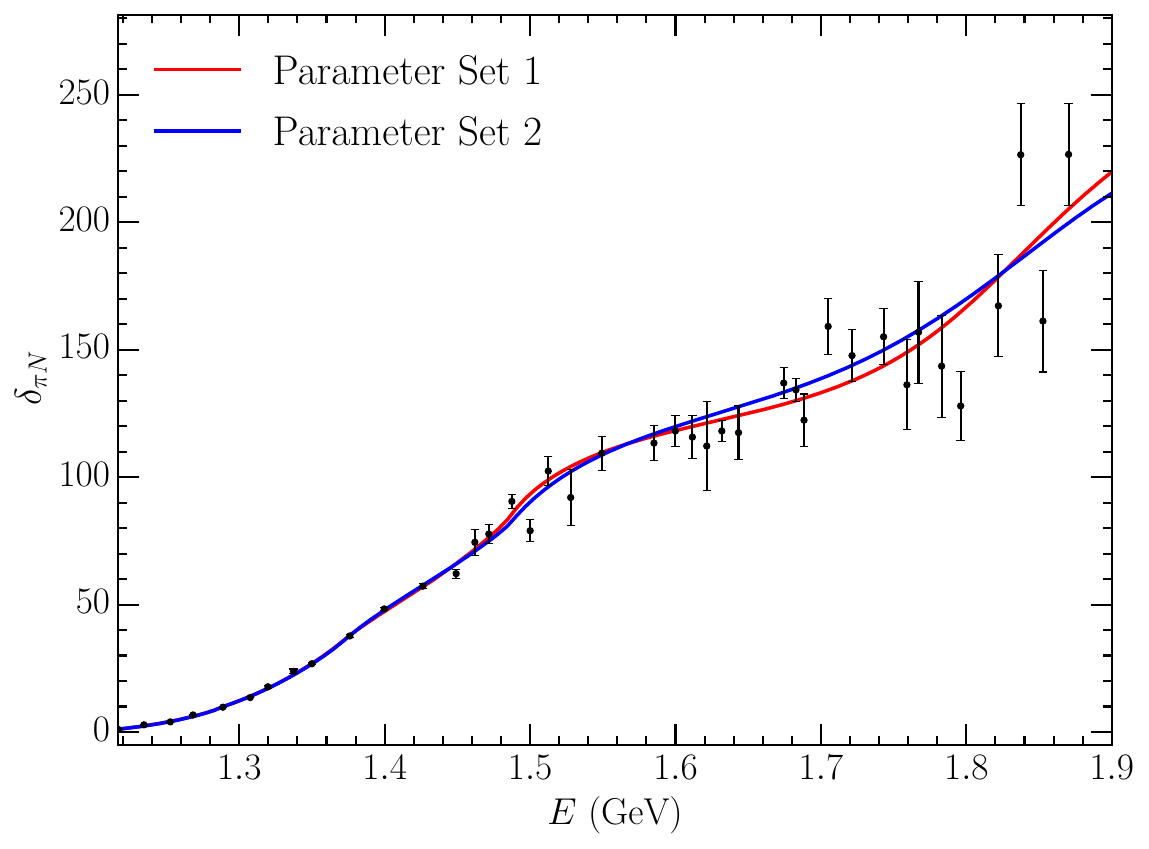}
    \end{minipage}
    \hfill
    \begin{minipage}[h]{.495\textwidth}
        \includegraphics[width=\textwidth]{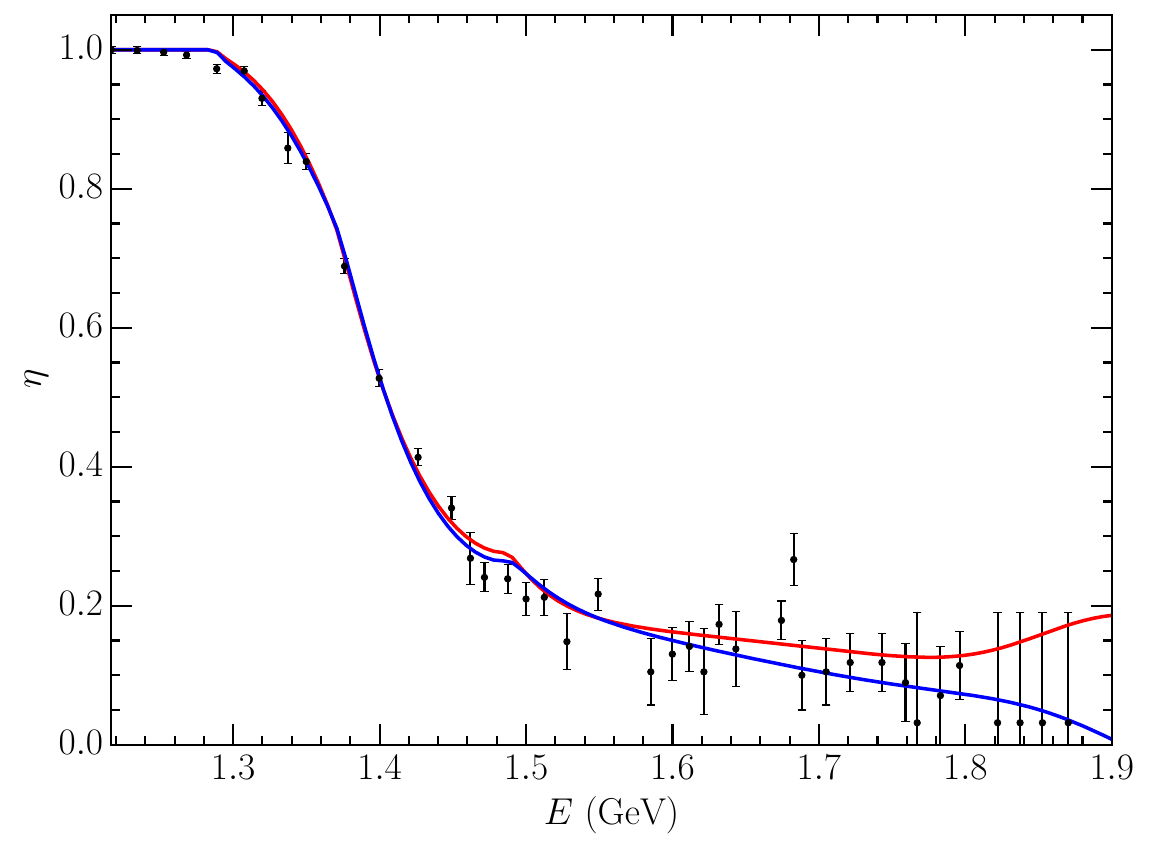}
    \end{minipage}
    \caption{Fits to experimental phase shift (left) and inelasticities (right) in $\pi N \to \pi N$ scattering for two parameter sets. The fits are performed over center of mass energies 1200 MeV to 1900 MeV obtained from SAID WI08 solution of $P_{ 1 1 }$ partial wave~\cite{GWU:2025ScatteringAmplitudes}.}
    \label{fig:fits_to_scattering}
\end{figure*}

The corresponding fits of these parameter sets to scattering observables are shown in Fig.~\ref{fig:fits_to_scattering}. While the HEFT coupling constants differ between the two sets, the calculated scattering observables agree well up to around 1800 MeV, beyond which the uncertainties increase.

The $\chi^{ 2 }$ values per degree of freedom ($\chi_{ \text{dof} }^{ 2 }$) for the fits to the experimental scattering observables are 159.24/(82$-$32) = 3.18 for parameter set 1 and 163.04/(82$-$32) = 3.26 for parameter set 2. Given the presence of a few outliers and the large uncertainties in the scattering results at higher center-of-mass energies, small (around 5\%) relative systematic uncertainties are added to the scattering phase shift results, consistent with previous HEFT analyses~\cite{Meissner:2000ChiSquared}. With this adjustment, the $\chi_{ \text{dof} }^{ 2 }$ values are reduced to 1.78 and 1.87 for Set 1 and Set 2, respectively. This adjustment does not meaningfully change the fitted HEFT parameters and is primarily introduced to provide a better illustration of the quality of fit.

These $\chi_{ \text{dof} }^{ 2 }$ values can be attributed to a couple of factors. We are missing threshold effects associated with the three-particle $\pi \pi N$ channel. While we do have the required $\sigma N$ and $\pi \Delta$ two-particle basis states associated with the resonant channels of $\pi \pi N$, we do not include their couplings to the three-particle channel. This channel may be particularly important near threshold and we approximate this physical effect by lowering the mass of the $\sigma$ meson when fitting experimental data. Secondly, while not dominant, channels such as $\rho N$, $\Lambda K$, and $\Sigma K$ have been neglected, despite their potential contributions to the dynamics.

It is worth noting that fitting exclusively to the scattering observables can yield a much smaller $\chi_{ \text{dof} }^{ 2 }$ value. However, such fits typically fail to describe the lattice QCD results accurately, highlighting their importance in the fitting process.

In Table~\ref{tab:pole_positions}, we present the pole positions of the Roper resonance and a second resonance near the $N(1880)$, along with their respective sheet locations for both parameter sets. For the Roper, we confirm that two associated poles are identified, one of which resides on the $uppu$-sheet and aligns more closely with the value provided by PDG~\cite{PDG:2022PoleLocations}. It is worth noting that in our model, depending on the mass chosen for the $\sigma$ in the effective $\sigma N$ channel, we may encounter a scenario where the $\pi \Delta$ channel is open while the $\sigma N$ channel remains closed. In such cases, an additional pole associated with the Roper may appear on the $uupp$-sheet. Indeed, we observe a pole in the $uupp$-sheet with a position of $E \approx 1.385 - 0.078i$ GeV for both parameter sets.

\begin{table}[b]
    \caption{Pole positions of the Roper resonance and the second higher mass resonance for both parameter sets. The location refers to the physical or unphysical Riemann sheets of complex energy for each channel, where the order of the sheets are $\left\lbrace s_{ \pi N }, s_{ \pi \Delta }, s_{ \eta N }, s_{ \sigma N } \right\rbrace$. We also quote the relevant PDG resonance positions from Ref.~\cite{PDG:2022PoleLocations}. } 
    \label{tab:pole_positions}
    \begin{ruledtabular}
        \begin{tabular}{cccc}
        Set & Location & Our value & PDG value\\[2pt]
        && (GeV) & (GeV) \\[2pt]
        \hline \\[-8pt]
        1 & $\left\lbrace u, p, p, u \right\rbrace$ & $1.367 - 0.117 i$ & $1.365(15) - 0.095(15) i$ \\
          & $\left\lbrace u, u, p, u \right\rbrace$ & $1.338 - 0.047 i$ \\
          & $\left\lbrace u, u, u, u \right\rbrace$ & $1.873 - 0.099 i$ & $1.860(40) - 0.115(25) i$ \\[2pt]
        \hline \\[-8pt]
        2 & $\left\lbrace u, p, p, u \right\rbrace$ & $1.363 - 0.120 i$ & $1.365(15) - 0.095(15) i$ \\
          & $\left\lbrace u, u, p, u \right\rbrace$ & $1.344 - 0.056 i$ \\
          & $\left\lbrace u, u, u, u \right\rbrace$ & $1.867 - 0.107 i$ & $1.860(40) - 0.115(25) i$
        \end{tabular}
    \end{ruledtabular}
\end{table}
\begin{figure*}
    \centering
    \begin{minipage}[h]{.495\textwidth}
        \includegraphics[width=\textwidth]{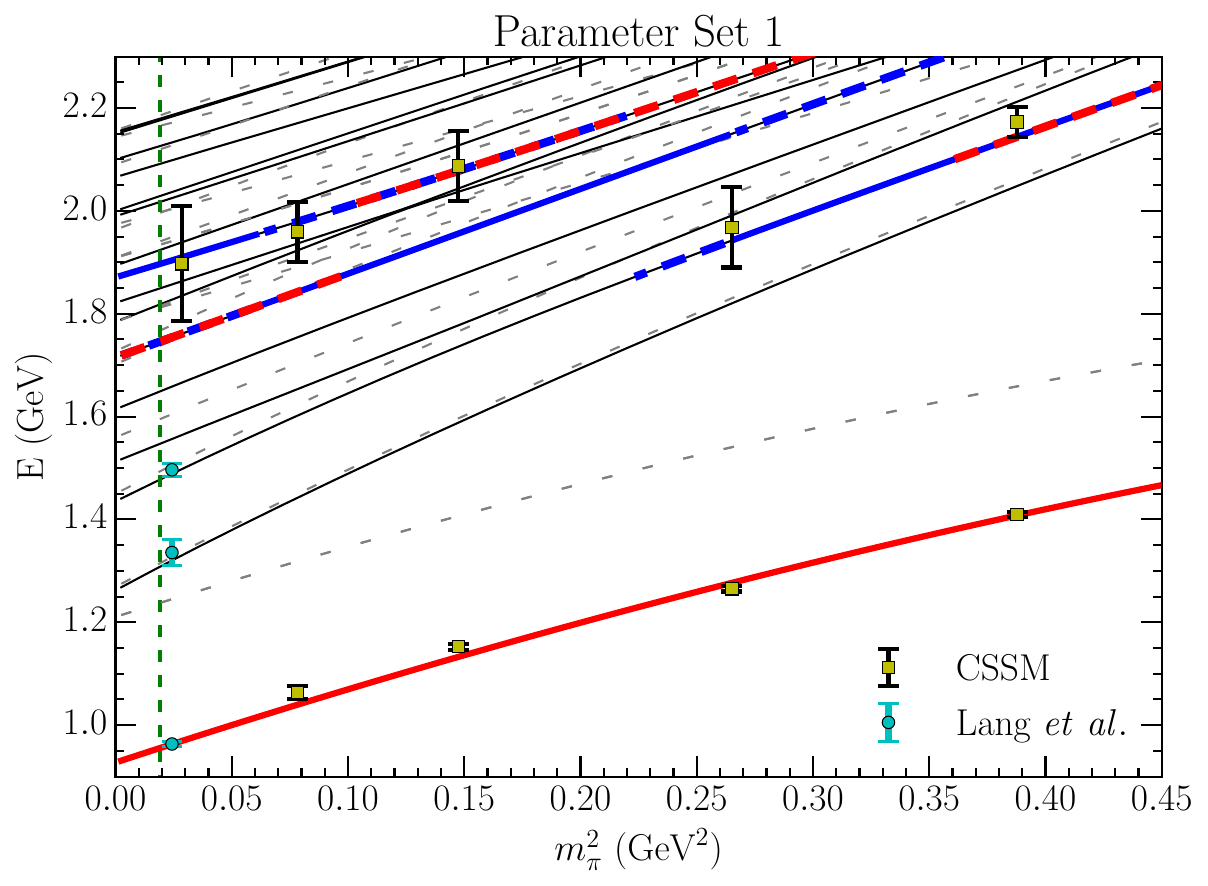}
    \end{minipage}
    \hfill
    \begin{minipage}[h]{.495\textwidth}
        \includegraphics[width=\textwidth]{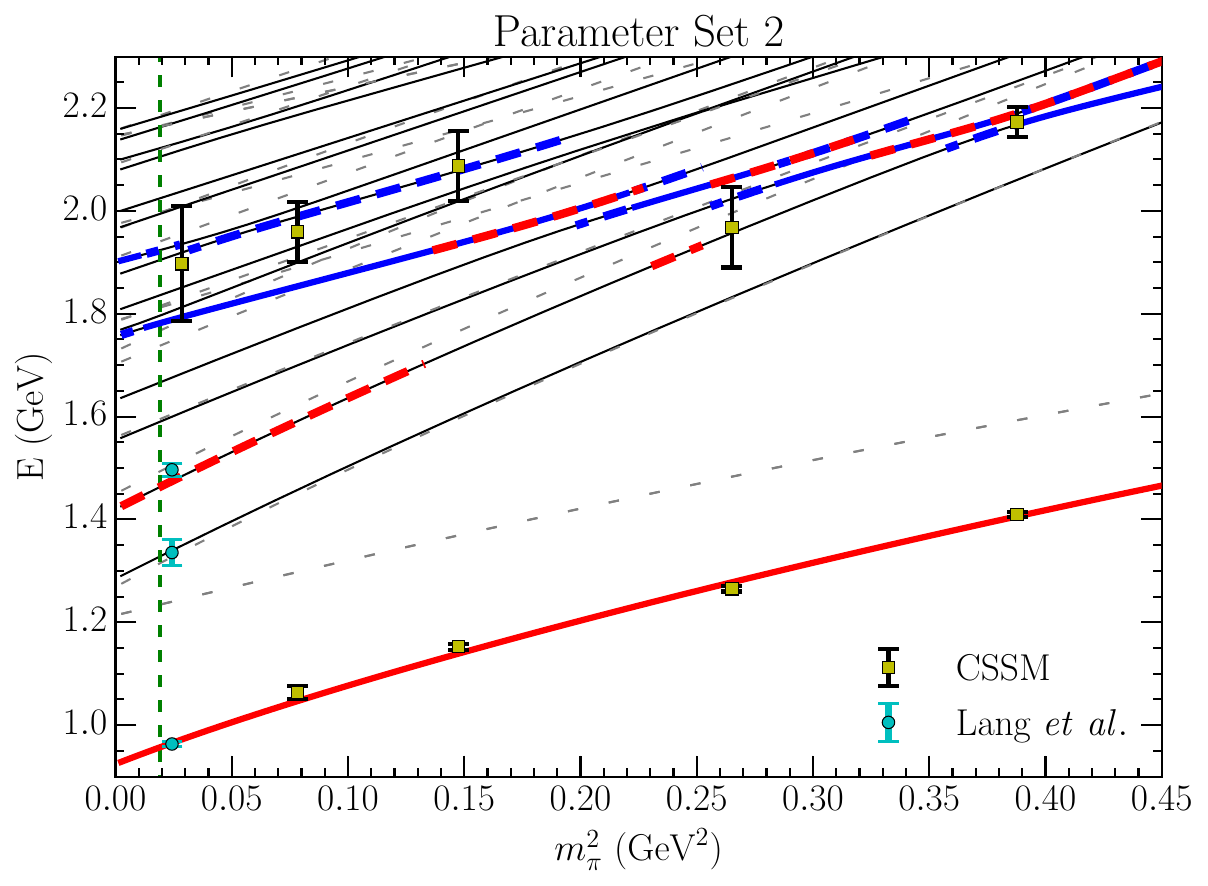}
    \end{minipage}
    \caption{Finite-volume energy spectrum as a function of $m_{ \pi }^{ 2 }$ for parameter set 1 (left) and 2 (right) at $L \approx 3$ fm. Lattice results used in the fit procedure are obtained from CSSM~\cite{Liu:2017Roper} and Lang \textit{et al}~\cite{Lang:2017LatticeResult}. The solid black lines represent the eigenvalues of the HEFT energy eigenstates $\ket{ E_{ i } }$, calculated via Eq.~\eqref{eq:eigenvalue_equation}. The dashed gray lines correspond to the energies of the noninteracting HEFT eigenstates. The red and blue lines (dashed lines) indicate the largest (second largest) contributions of the bare basis states to a given energy eigenstate, $\left| \left\langle N_{ 1 } \middle| E_{ i } \right\rangle \right|^{ 2 }$ and $\left| \left\langle N_{ 2 } \middle| E_{ i } \right\rangle \right|^{ 2 }$, respectively. The vertical green line marks the physical $m_{ \pi }^{ 2 }$ point.}
    \label{fig:finite_volume_spectrum_v_mpi2}
\end{figure*}

For the higher excited states of the nucleon we see in Table~\ref{tab:pole_positions} that we find only one pole, with a position closer to that quoted by the PDG for the $N(1880)$ ($E = 1.860(40) - 0.115(25)i$ GeV) than the $N(1710)$ ($E = 1.700(20) - 0.060(20)i$ GeV). This proximity to the $N(1880)$ is likely a consequence of the lattice QCD constraints used in our model. Specifically, since we rely on lattice results that feature a single three-quark state around 1.8–2.0 GeV, we are restricted to incorporating only two bare states (including the ground state nucleon). The pole positions obtained here suggest that a 2b4c system may be insufficient to fully describe both higher nucleon resonances.

According to the same CSSM lattice calculations, the second excited three-quark interpolated state sits approximately 60 MeV higher than the first excited state at the smallest pion mass. However, introducing a third bare basis state may not automatically resolve the reproduction of both the $N(1710)$ and $N(1880)$ resonance poles, as both excited states lie within the 1.8–2.0 GeV region. To fully capture the underlying physics, it appears necessary to incorporate additional coupled channels, constrained by more than just $\pi N \to \pi N$ scattering data. Nevertheless, the 2b4c system in this study remains fully sufficient for analyzing the Roper resonance.

% ---------------------------------------------------------------------------- %
% --------------------------- Finite-volume results -------------------------- %
% ---------------------------------------------------------------------------- %
\subsection{Finite-volume fits and results}

In the finite-volume, we use the value $L = 2.985$ fm, which is common to all the lattice QCD calculations. As noted earlier in this section, the slope parameters of the bare baryon mass expansions are tuned to ensure that HEFT energy eigenstates with significant bare state contributions align with three-quark interpolated lattice QCD results. The first bare state, $N_{ 1 }$, corresponds to the bare nucleon. Its mass, slope parameters, and series truncation naturally differ from those found in the renormalized nucleon in two-particle channels as summarized in Table~\ref{tab:two_particle_hadron_masses_and_slopes}.

Table~\ref{tab:bare_baryon_mass_and_slope} lists the mass expansion parameters for the bare baryon states in both parameter sets. The slope parameters of $N_{ 1 }$ were tuned to the ground state nucleon lattice results from CSSM~\cite{Liu:2017Roper}, except for the result near the physical pion mass, which was taken from Lang \textit{et al.}~\cite{Lang:2017LatticeResult}. The slope parameters for $N_{ 2 }$ are tuned to the first excited nucleon state from CSSM~\cite{Liu:2017Roper}. Since the bare mass at the physical point corresponds to the values listed in Tables~\ref{tab:parameter_set_1} and \ref{tab:parameter_set_2}, the evaluation point is fixed at $m_{ \pi } \approx 0.139$ GeV. Notably, at least for the bare state $N_{ 1 }$, the inclusion of the next-to-leading order slope parameter, $\alpha_{ B_{ 0 }, 2 }$, is necessary to capture the curvature of nucleon mass as a function of $m_{ \pi }^{ 2 }$. In Fig.~\ref{fig:finite_volume_spectrum_v_mpi2}, we present the finite-volume energy spectra as a function of $m_{ \pi }^{ 2 }$ for both parameter sets. 

Although both parameter sets exhibit similar overall features, a key distinction lies in the behavior of the sixth energy eigenstate (black line). In Set 1, the energy eigenstates remain largely parallel, whereas in Set 2, this behavior is absent. From Fig.~\ref{fig:finite_volume_spectrum_v_mpi2}, the HEFT solutions suggest that the mass of the radial excitation observed in lattice QCD should sit a little lower than the three lightest pion mass lattice QCD eigenstates at 1.9 to 2.0 GeV. However, for the two heaviest pion masses, HEFT describes the lattice results well. To gain further insight, we turn to an eigenvector analysis, which reveals the contributions of various basis states to each energy eigenstate.

\begin{table}[b]
    \caption{Bare baryon mass and slope parameters of mass expansion Eq.~\eqref{eq:hadron_mass_expansion} for both parameter sets. The bare mass is taken from Tables~\ref{tab:parameter_set_1} and \ref{tab:parameter_set_2}, and the evaluation point is set to the physical point $m_{ \pi } \approx 0.139$ GeV.} 
    \label{tab:bare_baryon_mass_and_slope}
    \begin{ruledtabular}
        \begin{tabular}{ccccc}
        Set & $B_{ 0 }$ & $\left. m_{ B_{ 0 } } \right|_{ \text{e.p.} }$ & $\alpha_{ B_{ 0 }, 1 }$ & $\alpha_{ B_{ 0 }, 2 }$ \\[2pt]
        && (GeV) & (GeV$^{ -1 }$) & (GeV$^{ -3 }$) \\[2pt]
        \hline \\[-8pt]
        1 & $N_{ 1 }$ & 1.237 & 1.438 & -0.727 \\
          & $N_{ 2 }$ & 1.934 & 1.304 &  0.468 \\[2pt]
        \hline \\[-8pt]
        2 & $N_{ 1 }$ & 1.234 & 1.101 & -0.316 \\
          & $N_{ 2 }$ & 2.164 & 0.960 & -0.073
        \end{tabular}
    \end{ruledtabular}
\end{table}
\begin{figure*}
    \centering
    \includegraphics[width=\textwidth]{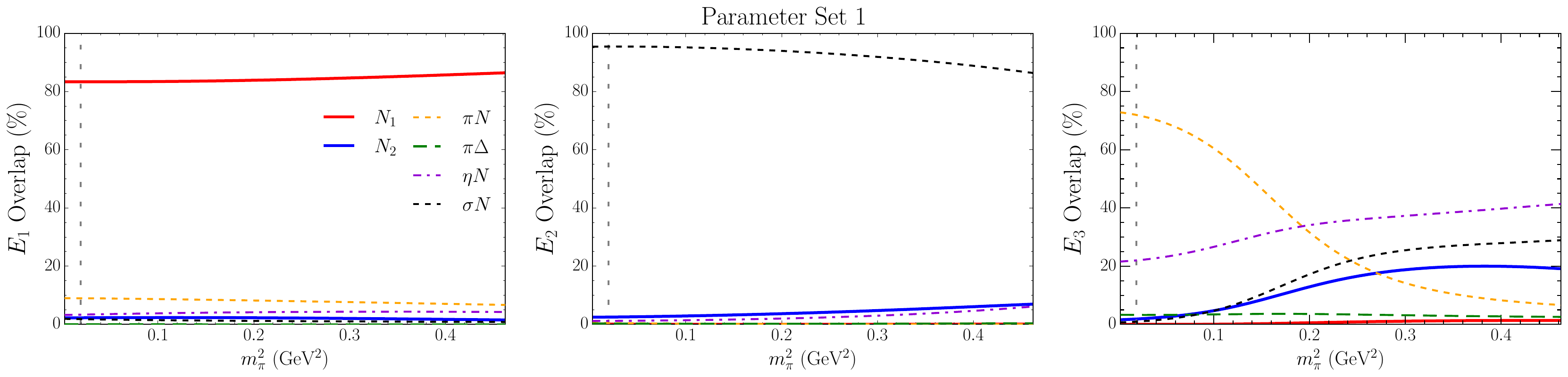}
    \includegraphics[width=\textwidth]{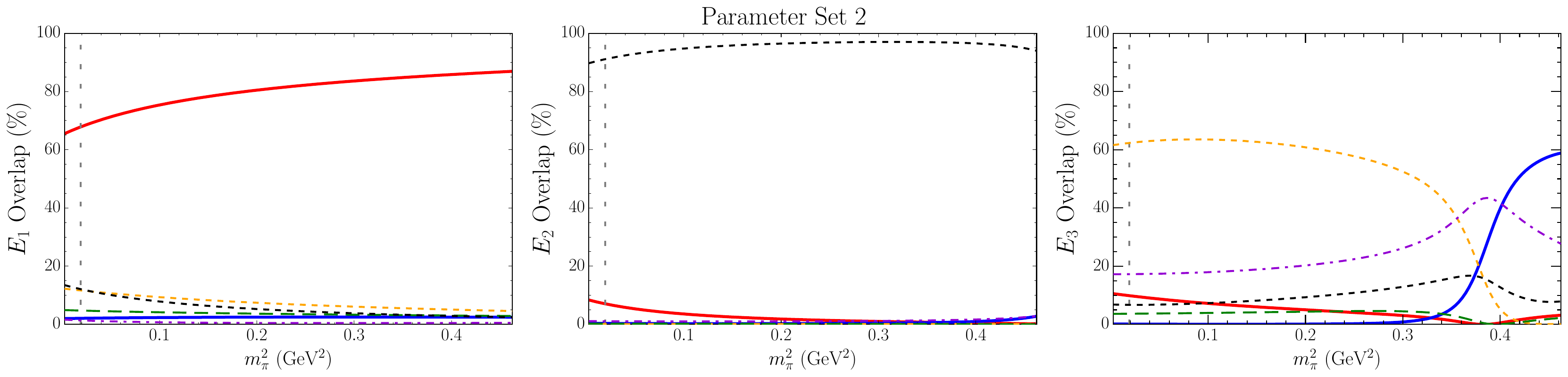}
    \caption{Overlap for each basis state $\ket{ \mathcal{ B }\, }$ to each energy eigenstate $\left| \left\langle \mathcal{ B }\, \middle| E_{ i } \right\rangle \right|^{ 2 }$, as a function of $m_{ \pi }^{ 2 }$ for both parameter sets. For the two-particle channels, the discrete momenta are summed as $\sum_{ n } \left| \left\langle \alpha( k_{ n } ) \middle| E_{ i } \right\rangle \right|^{ 2 }$.}
    \label{fig:overlap_v_mpi2_1}
\end{figure*}

\begin{figure*}
    \centering
    \includegraphics[width=\textwidth]{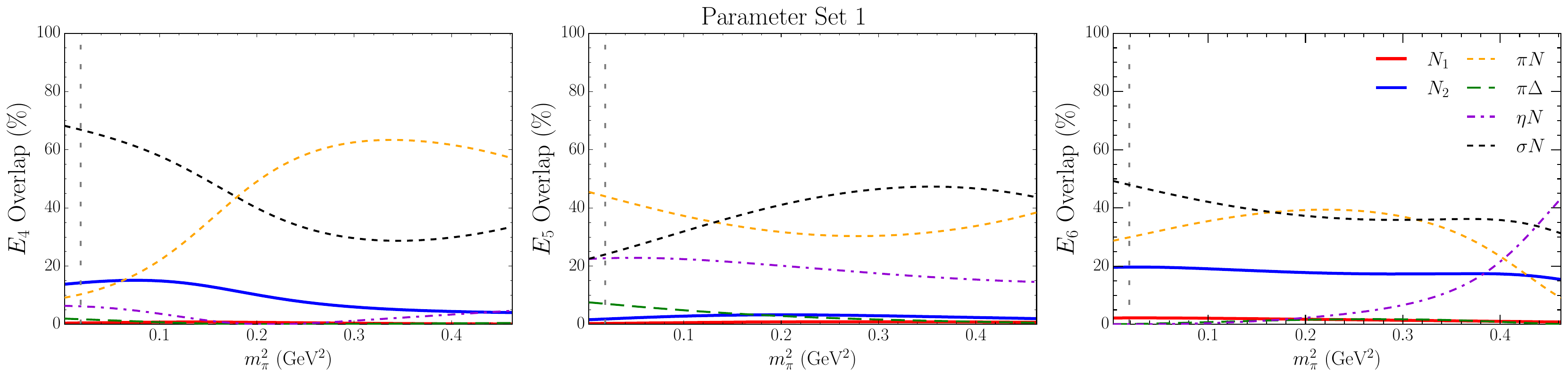}
    \includegraphics[width=\textwidth]{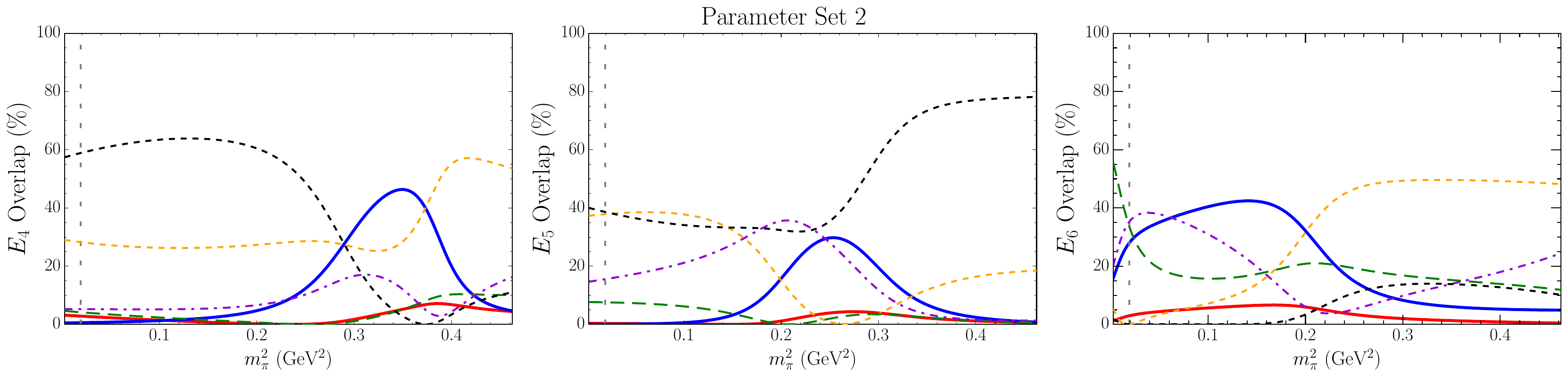}
    \caption{Same figures as Fig.~\ref{fig:overlap_v_mpi2_1} but for energy levels $E_{ 4 }$, $E_{ 5 }$, and $E_{ 6 }$.}
    \label{fig:overlap_v_mpi2_2}
\end{figure*}

\begin{figure*}
    \centering
    \includegraphics[width=\textwidth]{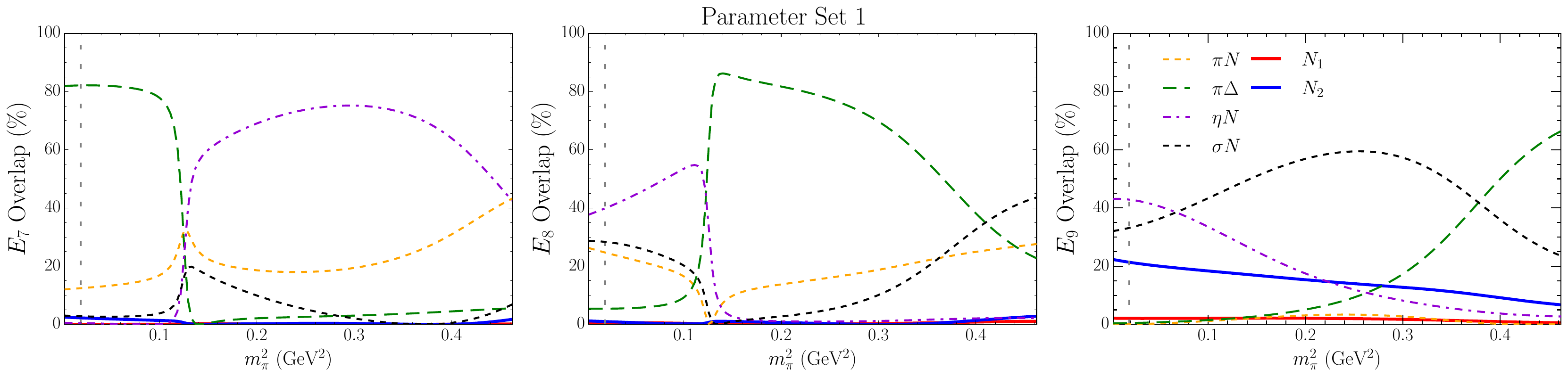}
    \includegraphics[width=\textwidth]{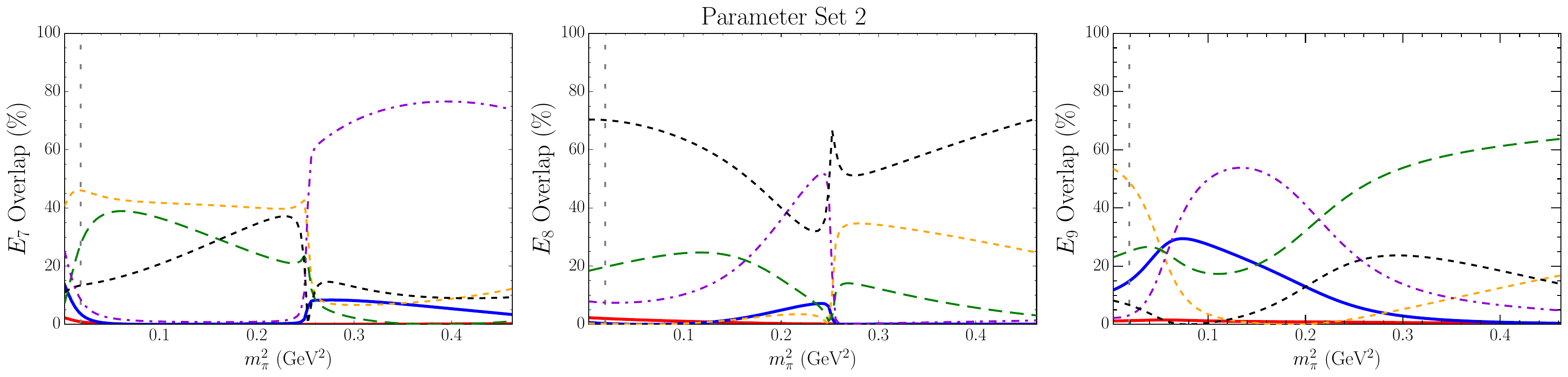}
    \caption{Same figures as Figs.~\ref{fig:overlap_v_mpi2_1} and \ref{fig:overlap_v_mpi2_2} but for energy levels $E_{ 7 }$, $E_{ 8 }$, and $E_{ 9 }$.}
    \label{fig:overlap_v_mpi2_3}
\end{figure*}

In Figs.~\ref{fig:overlap_v_mpi2_1}, \ref{fig:overlap_v_mpi2_2}, and \ref{fig:overlap_v_mpi2_3}, we present the relative contribution, expressed as a percentage, of each Hamiltonian basis state
\begin{equation*}
    \ket{ \mathcal{ B }\, } \in \left\lbrace \, \ket{ N_{ 1 } }, \ket{ N_{ 2 } }, \ket{ \pi N( k_{ 0 } ) }, \cdots, \ket{ \sigma N( k_{ \text{max} } )\, } \right\rbrace \, ,
\end{equation*}
to each energy eigenstate $\ket{ E_{ i } }$. These contributions are calculated via the overlap $\left| \left\langle \mathcal{ B }\, \middle| E_{ i } \right\rangle \right|^{ 2 }$, where each energy eigenvector is expressed analogously to Eq.~\eqref{eq:energy_eigenket}. Instead of examining the contribution from individual two-particle channel basis states $\ket{ \alpha( k_{ n } ) }$ at a specific momentum $k_{ n }$, we consider the total contribution from each channel, \textit{i.e.}, $\sum_{ n } \left| \left\langle \alpha( k_{ n } ) \middle| E_{ i } \right\rangle \right|^{ 2 }$.


A comparison of the two parameter sets reveals some differences in the overlaps. However, when focusing on the first three energy levels at the physical point (indicated by the vertical gray dashed line), the main features remain consistent.

For the first energy level, the energy eigenstate is predominantly composed of the bare baryon state $\ket{ N_{ 1 } }$, with small contributions from $\pi N$ and $\sigma N$ channels. This is the prototypical picture of the ground state nucleon: a three-quark core with some pion dressing. The second energy level is dominated by the $\sigma N$ channel, with a small contribution from $\ket{ N_{ 1 } }$. This is in accord with the lattice correlation matrix analysis of Ref.~\cite{Lang:2017LatticeResult}. The third energy level is primarily associated with $\pi N$ channel, again in accord with the lattice analysis~\cite{Lang:2017LatticeResult}. 

In infinite-volume, the Roper resonance is located between the second and third energy levels. It is unsurprising to see that there is only a small contribution from the bare states, as no three-quark interpolated states in lattice QCD are found in this energy region. In fact, for larger lattice extents ($L > 3.0$ fm) at the physical point, a similar structure emerges: HEFT energy eigenstates around the Roper resonance are dominated by the $\pi N$ and $\sigma N$ channels. Moreover, basis states like the $\eta N$ or $N_{ 1 }$, start to contribute less, with their maximum combined contribution being approximately 10\%. This behavior is consistent across both parameter sets and likely reflects a general feature of the underlying physics.

As illustrated in Figs.~\ref{fig:overlap_v_mpi2_2} and \ref{fig:overlap_v_mpi2_3}, beyond the third energy level the degree of model dependence becomes more pronounced. Near the physical point, the two nucleon resonances, $N(1710)$ and $N(1880)$, are in the neighborhood of the sixth and ninth energy eigenstates in Figs.~\ref{fig:overlap_v_mpi2_2} and \ref{fig:overlap_v_mpi2_3} (rightmost column) for both parameter sets. The only commonality between the sets is that these resonances exhibit roughly a 20\% contribution from the $N_{ 2 }$ bare state, with a negligible contribution from $N_{ 1 }$. The remaining contributions stem from two-particle channels. However, the dominant channels differ significantly between the two parameter sets. For instance, in Set 1, the dominant two-particle channels are $\sigma N$ and $\pi N$, whereas in Set 2, they are $\pi \Delta$ and $\eta N$. This highlights a case where improved experimental data on inelastic channels would be highly beneficial. Similarly, precision scattering state positions from lattice QCD would be of interest.

Beyond the physical point, the contribution of the $N_{ 2 }$ bare state is more pronounced in Set 2 compared to Set 1, where it exhibits peaks in contributions to specific energy levels, as seen in energy levels $E_{ 3 }$–$E_{ 6 }$ and $E_{ 9 }$ of Figs.~\ref{fig:overlap_v_mpi2_1}, \ref{fig:overlap_v_mpi2_2}, and \ref{fig:overlap_v_mpi2_3}. This is further reflected in Fig.~\ref{fig:finite_volume_spectrum_v_mpi2} (right panel), where the dominant contributions from the $N_{ 2 }$ state span multiple energy levels. In contrast, the $N_{ 2 }$ contributions in Set 1 are more diffuse, with a relatively smooth $m_{ \pi }^{ 2 }$ dependence and contributions that are more evenly distributed across energy levels $E_{ 3 }$, $E_{ 6 }$, and $E_{ 9 }$. 

We now return to the previous discussion, where both parameter sets exhibit a significant $N_{ 2 }$ bare state contribution at energies slightly below the excited three-quark interpolated lattice states at the three lightest pion masses. The excited states at these pion masses correspond to the ninth HEFT eigenstate, $E_{ 9 }$, for both parameter sets. In considering the fits, we attempted to tune the pion mass dependence of $N_{ 2 }$ to effectively maximize the overlap $\left| \left\langle N_{ 2 } \middle| E_{ 9 } \right\rangle \right|^{ 2 }$ at the three lightest pion masses. Despite this, we observe at most an approximate 30\% bare state contribution to $E_{ 9 }$. Hence, examining both the basis state decompositions and the positioning of the dominant second bare basis state suggests that the excited state seen on the lattice, while including a bare component, exhibits a rather a more complex structure.

\begin{figure*}
    \centering
    \begin{minipage}[h]{.495\textwidth}
        \includegraphics[width=\textwidth]{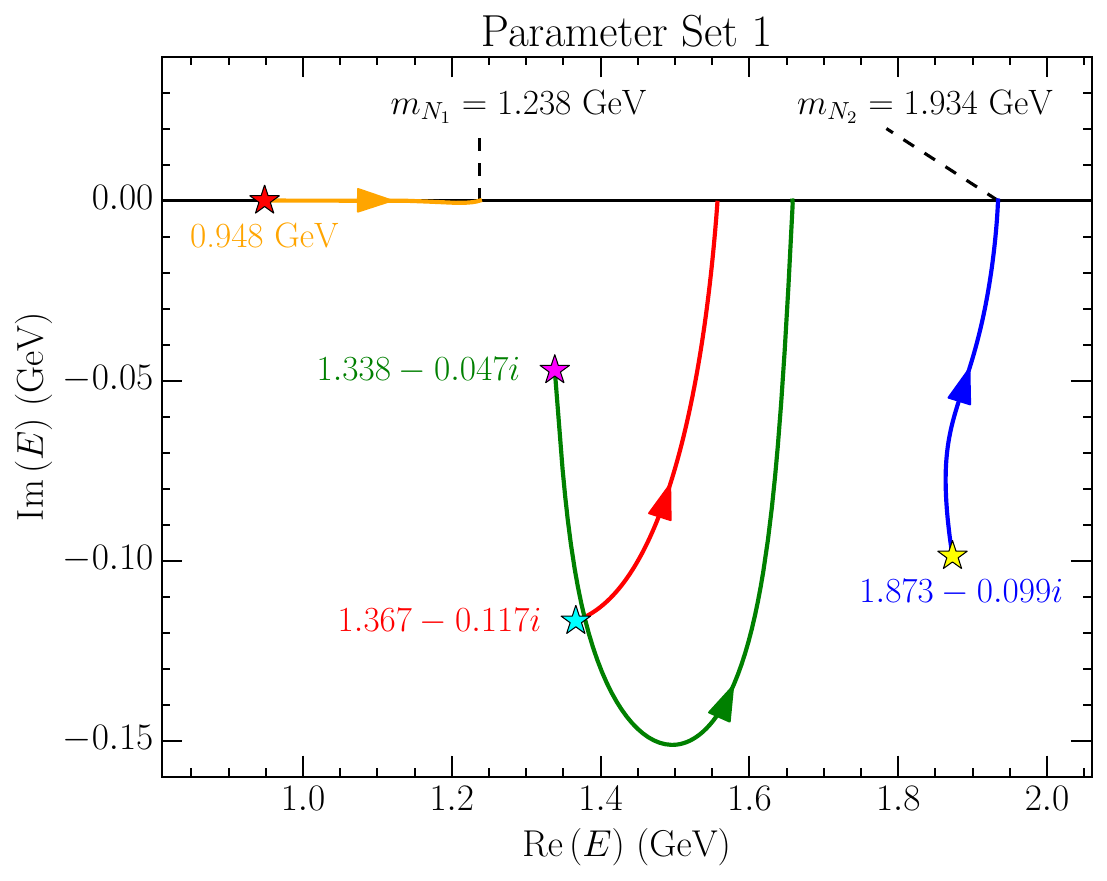}
    \end{minipage}
    \hfill
    \begin{minipage}[h]{.495\textwidth}
        \includegraphics[width=\textwidth]{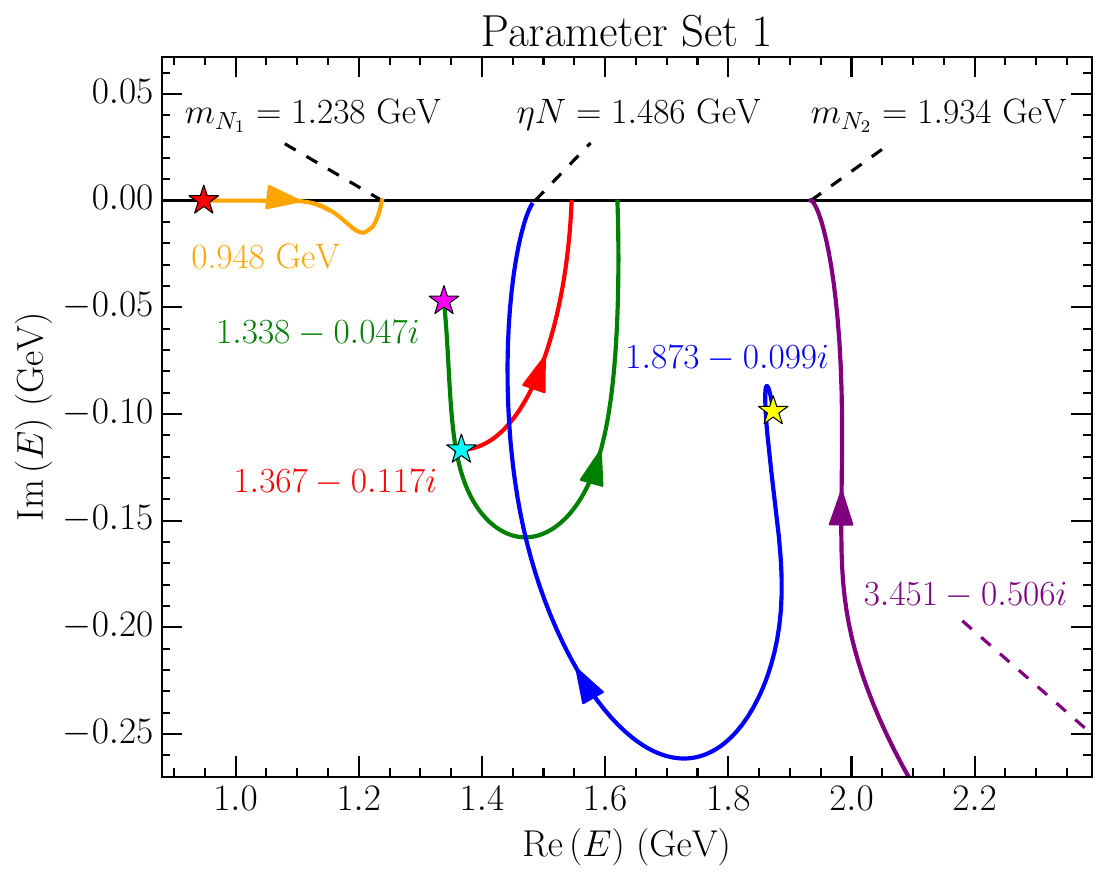}
    \end{minipage}
    \caption{Pole trajectories of the poles listed in Table~\ref{tab:pole_positions} for Set 1. The left panel shows the trajectories when all the coupling constants of HEFT are linearly turned off from 100\% to 0\%, while the right panel shows the trajectories when the regulator parameters are similarly reduced. Arrows indicate the direction of the pole movements. The stars denote the starting positions of the poles, except for the pole associated with the purple trajectory, which begins outside the panel at $3.451 - 0.506 i$ GeV.}
    \label{fig:pole_trajectory}
\end{figure*}

These lattice QCD results at 1.9 to 2.0 GeV are bracketed by two HEFT states both of which have a large $N_{ 2 }$ bare basis state component. As the three-quark operators considered in the lattice QCD analysis will excite both eigenstates, one relies on Euclidean time evolution to suppress the excited state. However, extracting states at this energy is challenging and ample Euclidean time evolution is not available on the lattice. As such, we conclude that the three excited states reported from the lattice are most likely superpositions of the two HEFT states dominated by $N_{ 2 }$ and the mass reported is thus a weighted average of these two energy levels.

There is an indication of a cleanly resolved three-quark excited state when considering the heaviest pion mass at $m_{ \pi }^{ 2 } \approx 0.4$ GeV$^{ 2 }$, with a dominant $N_{ 2 }$ contribution aligned with it in both parameter sets. The eigenvector decomposition (see the overlap of $E_{ 3 }$ in Fig.~\ref{fig:overlap_v_mpi2_1}) shows strong contributions from $\eta N$ and $\sigma N$ channels, indicating the possibility that the excited state may consist of a three-quark core with pion and $\eta$ dressing.

Since the second excited state in the 1.9 GeV region of the lattice calculations is in close proximity to the first excited state, a study using a 3b4c system could potentially help resolve the structure of both nucleon resonance poles.

% ---------------------------------------------------------------------------- %
% ----------------------------- Pole trajectories ---------------------------- %
% ---------------------------------------------------------------------------- %
\subsection{Pole trajectories}

By combining experimental and lattice QCD results, we observe that the Roper resonance is predominantly a dynamically generated state, with a potential weak interaction with a bare nucleon state. On the other hand, Ref.~\cite{Suzuki:2010PoleTrajectory} proposes that the Roper (or the poles associated with it) and the $N(1710)$ both evolve from a single bare state through their couplings with the $\pi N$, $\eta N$, and $\pi \pi N$ channels. We now consider the evolution of the resonant states, concluding that the method used in Ref.~\cite{Suzuki:2010PoleTrajectory} to identify the origin of any state may not yield a physical interpetation.

An approach to uncovering the origin of a resonance is to trace the trajectory of the pole by ``turning on" or ``turning off" the coupling constants of a model. For example, in Ref.~\cite{Suzuki:2010PoleTrajectory}, the coupling constants of the model were gradually increased from 0\% to 100\%, while the $\pi \Delta$ self-energy term was held fixed at 0\%. The two poles associated with the Roper emerge only after the $\pi \Delta$ self-energy term is activated.

Here, we trace the trajectories of all the poles listed in Table~\ref{tab:pole_positions} (along with an additional pole described later) by linearly turning off either {\em all} the coupling constants (method 1) or {\em all} the regulator parameters (method 2) in the HEFT model. Both methods yield the same initial and final pole positions by fully turning on or off any channel and/or bare state contributions. However, the paths taken by the poles can differ significantly. In Fig.~\ref{fig:pole_trajectory}, we show how the poles move in the complex $E$-plane for Set 1. The left panel displays the trajectories when the coupling constants are turned off, while the right panel shows the effect of turning off the regulator parameters. 

\begin{figure*}
    \centering
    \begin{minipage}[h]{.48\textwidth}
        \includegraphics[width=\textwidth]{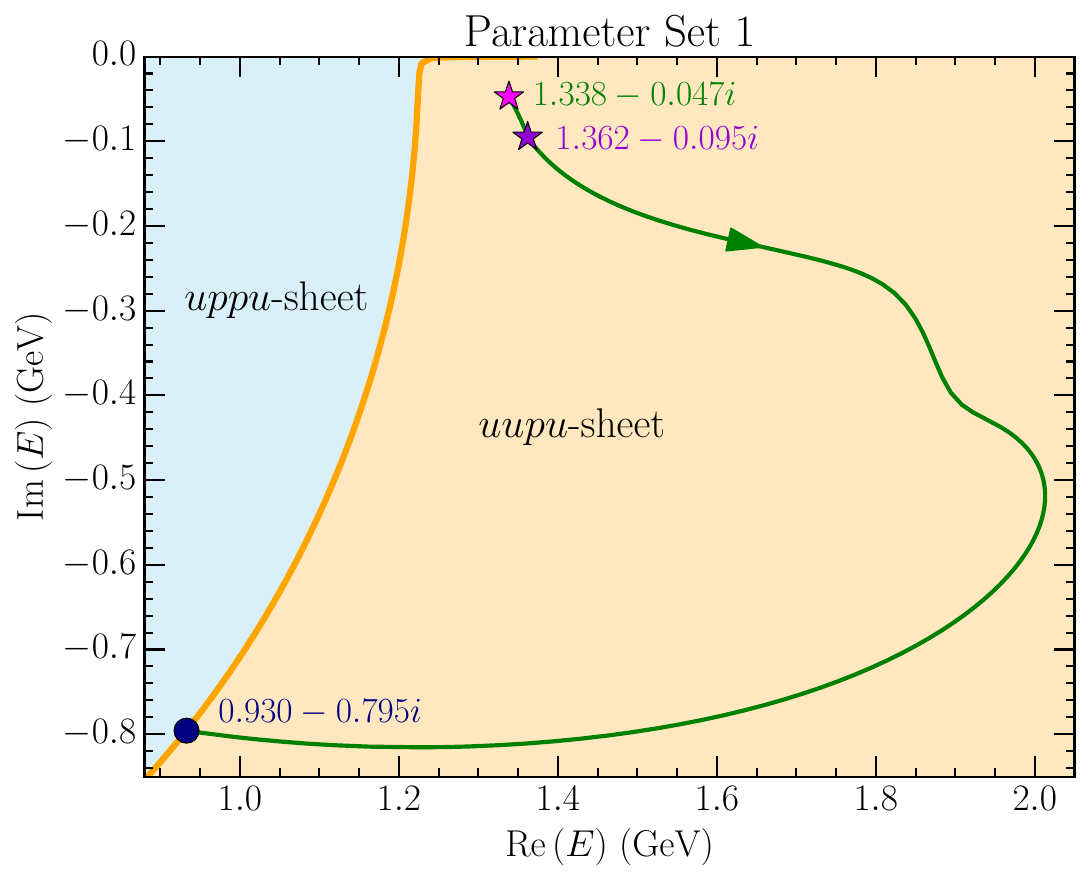}
    \end{minipage}
    \hfill
    \begin{minipage}[h]{.503\textwidth}
        \includegraphics[width=\textwidth]{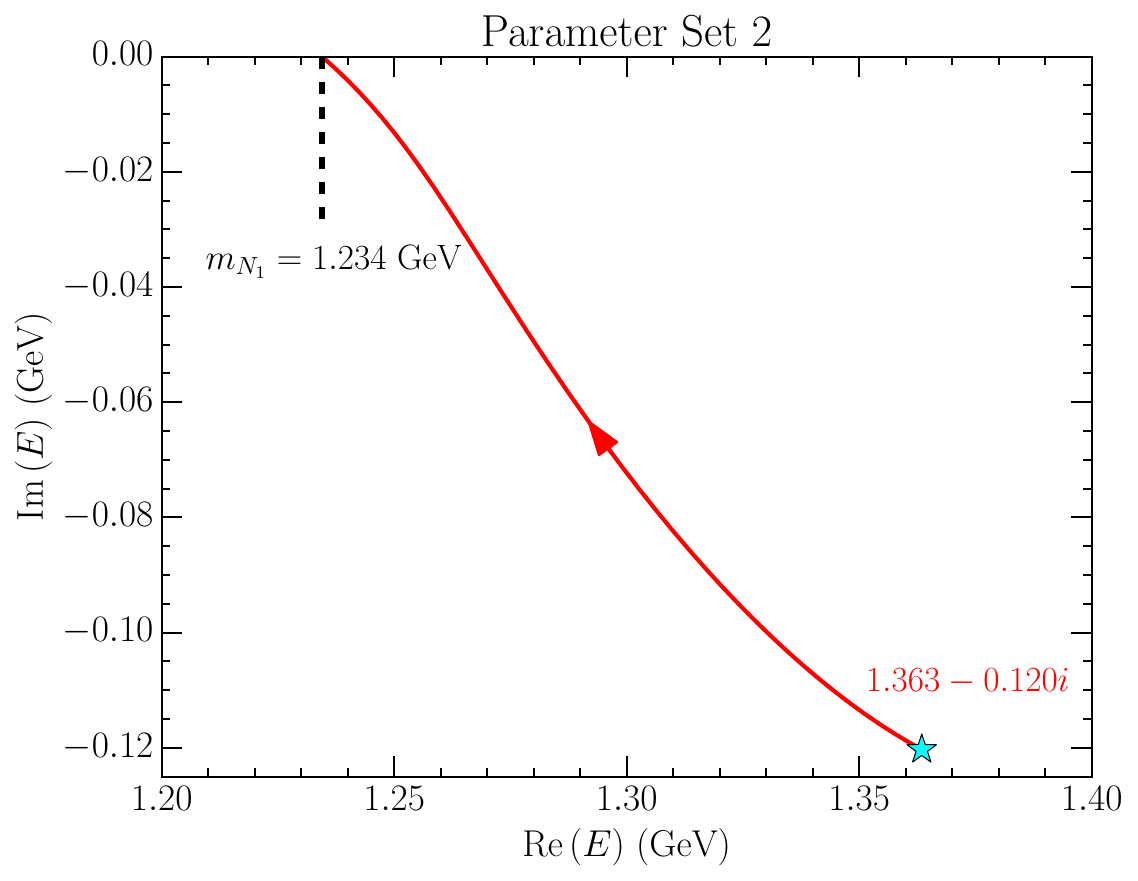}
    \end{minipage}
    \caption{The pole trajectories of poles associated with the Roper resonance when turning off the 1-to-2 couplings first, then the 2-to-2 couplings, for both parameter sets -- see the text for a detailed discussion. The orange line in the left figure shows the contour line of $\pi \Delta$ channel in complex $E$-plane. In the complex $q$-plane, the contour effectively runs almost directly down the imaginary axis, where $\theta_{ \pi \Delta } \approx -90^{ \circ }$.}
    \label{fig:pole_trajectory_2}
\end{figure*}

We observe that, despite minor discrepancies between the two methods, the nucleon and the two poles associated with the Roper follow similar trajectories. As the parameters are turned off, the two Roper poles move toward the real axis in approximately the same locations for both methods. Notably, they reach the real axis before the parameters are fully turned off — only about a third of the way through the process — at points that do not align with any two-particle thresholds or the bare masses. This contrasts with the pole trajectories in Ref.~\cite{Suzuki:2010PoleTrajectory}, where the Roper poles unambiguously emerge from a bare state.

The pole associated with the $N(1880)$ follows a disparate trajectory. In method 1, as the couplings are fully turned off, the pole moves to the real axis, where its energy aligns with the mass of the second bare state, $m_{ N_{ 2 } } = 1.934$ GeV. In method 2, it also reaches the real axis — albeit with some detour — at the $\eta N$ threshold of 1.486 GeV. The most notable difference, however, is that in method 2, a completely different pole (unassociated with any known resonances), initially located at $3.451 - 0.506 i$ GeV, takes the place of the $N(1880)$ pole and terminates at $m_{ N_{ 2 } }$.

Performing the same analysis for Set 2 reveals a clear model dependence, where the pole termination points differ significantly between the two parameter sets. For method 1, the trajectories of the nucleon and the $N(1880)$ pole are similar to those in Set 1. However, the Roper poles wander around the complex $E$-plane, eventually crossing a contour line and ending up on different Riemann sheets. In method 2, the two Roper poles terminate on the real axis, both at the $\sigma N$ threshold of 1.289 GeV. The $N(1880)$ pole, similar to Set 1, terminates at the $\eta N$ threshold. Again, in this case, another pole terminates at the mass of the second bare state, $m_{ N_{ 2 } } = 2.164$ GeV.

It may seem intuitive to determine the origin of a resonance by tracing the initial and final positions of the corresponding pole; however, such an inference cannot and should not be made without a well-defined, systematic procedure — if one even exists — for turning the parameters on or off. This issue is particularly evident in the case of the $N(1880)$ pole, where different methods result in the pole terminating at different positions across both parameter sets. Without such a definitive procedure, the trajectories and termination points of the poles are not uniquely determined. In other words, only the initial pole positions and those at the bare masses — when all coupling and regulator parameters are turned off — are fixed.

Given that method 1 is the simpler of the two, we apply Occam's razor and build on the idea that only the initial data-constrained positions are fixed. In Fig.~\ref{fig:pole_trajectory_2}, we illustrate how the poles associated with the Roper evolve for both parameter sets using method 1, but in a stepwise manner: first, we linearly turn off the 1-to-2 couplings, followed by the 2-to-2 couplings.

For Set 1, recall that the system is based on a 0b4c system, where pure 2-to-2 poles are modified in the full 2b4c system. However, in the limit where the 1-to-2 couplings are small, the full system recovers the pure 2-to-2 pole. This behavior is demonstrated in the left panel of Fig.~\ref{fig:pole_trajectory_2}, where we start with the Roper pole on the $uupu$-sheet at $E = 1.338 - 0.047i$ GeV. As the 1-to-2 couplings are gradually reduced (but not fully), the pole approaches the 2-to-2 pole at $E = 1.362 - 0.095i$ GeV. Coincidentally, this pole position coincides with the Roper listed by the PDG~\cite{PDG:2022PoleLocations}, at $E = 1.365(15) - 0.095(15)i$ GeV. From here, we proceed to turn off the 2-to-2 couplings. The pole then travels — again, with some detour — toward a position associated with the 2-to-2 $\pi \Delta$ regulator pole, located at $E = 0.930 - 0.795i$ GeV.

To clarify, regulator poles which approximate the physical left-hand cut structure of the scattering amplitude, arise from the dipole form of the regulators, with poles occurring at values of complex momentum $q = i \Lambda$ for each regulator. In the figure, the curved orange line demarcates the different Riemann sheets (the $\pi \Delta$  channel boundary), and the regulator pole lies on top of this boundary as the complex momentum is rotated $q \to q\, e^{ i \theta_{ \pi \Delta } }$, where $\theta_{ \pi \Delta } \approx -90^{ \circ }$. Therefore, $E = 0.930 - 0.795i$ GeV corresponds to $q = i \Lambda_{ \pi \Delta } = 0.807i$ GeV, matching the value presented in Table~\ref{tab:parameter_set_2}.

For Set 2, we find no 2-to-2 poles near the Roper, consistent with the behavior of the base 1b4c system. In this case, as the 1-to-2 couplings are turned off, the Roper pole on the $uupu$-sheet moves toward the mass of the bare nucleon $N_{ 1 }$, where $m_{ N_{ 1 } } = 1.234$ GeV, as shown in the right panel of Fig.~\ref{fig:pole_trajectory_2}. Once the pole reaches this position, further turning off the 2-to-2 couplings has no impact on its location. 

In both parameter sets, we observe that the trajectories of the poles depend on how the parameters are varied. The path a pole follows, as well as its termination point, is not unique. Without a definitive “correct” way to vary the parameters, we can only reasonably conclude that these pole trajectories do not meaningfully represent the true physical origin of the resonance.

% ---------------------------------------------------------------------------- %
% -------------------------------- Conclusions ------------------------------- %
% ---------------------------------------------------------------------------- %
\section{Conclusions}\label{section:conclusion}

We have presented a detailed study of the nature of the Roper $N(1440)$ resonance using the Hamiltonian Effective Field Theory formalism. It is the first analysis to fully incorporate two bare basis states and their interactions in the positive parity nucleon spectrum. We generated two valid HEFT parameter sets, each incorporating two bare states and four coupled channels, which sufficiently reproduce experimental phase shifts, inelasticities, and resonance positions while also aligning with three-quark interpolated states from lattice QCD.

Based on the HEFT eigenvector analysis with these two parameter sets, the Roper resonance can be understood as a dynamically generated state with negligible contributions from pure bare states, in concordance with previous HEFT analyses of Refs.~\cite{Wu:2017qve,Liu:2017Roper}. The interpretation of the next two excited nucleon states, $N(1710)$ and $N(1880)$, remains challenging. There is some indication that either the $N(1710)$ or the $N(1880)$ consists of a three-quark core with pion and $\eta$ dressings. While our pole position favors the $N(1880)$ resonance, the analysis omits a description of the $N(1710)$. This suggests a HEFT analysis incorporating three bare states and meson-baryon channels incorporating hyperons may be required to complete the nucleon resonance spectrum in this energy regime. However, to pursue this, precision spectra from the lattice incorporating a comprehensive set of two-particle interpolating fields (in the first instance) is required.

We have also explored the origins of the resonant states by tracing the resonance poles as couplings and regulators are turned off. Our analysis demonstrates that this method does not provide a meaningful physical understanding of the origin of resonances. The trajectories of the poles and their termination points are not uniquely determined. Rather, their paths depend on the details of how the parameters are varied, and the complex interplay of self-energies can give rise to many different interpretations.

% ---------------------------------------------------------------------------- %
% ------------------------------- Bibliography ------------------------------- %
% ---------------------------------------------------------------------------- %
\section*{Acknowledgements}
This work was supported by the University of Adelaide and by the Australian Research Council through Discovery Projects DP210103706 (DBL) and DP230101791 (AWT).  

\bibliography{bibliography}

\end{document}